\documentclass[rmp]{revtex4}
\usepackage{amsmath}
\usepackage{amsfonts}
\usepackage{amssymb}
\usepackage{graphicx}
\usepackage{color}

\renewcommand{\a}{\alpha}

\newcommand{\g}{\gamma}
\renewcommand{\k}{\kappa}
\renewcommand{\l}{\lambda}
\newcommand{\s}{\sigma}
\newcommand{\sq}{\sigma^2}
\newcommand{\E}{\mathbb{E}}

\renewcommand{\P}{\mathbb{P}}

\newcommand{\N}{\mathcal{N}}
\renewcommand{\L}{\Lambda}
\newcommand{\ud}{\,\mathrm{d}}

\newcommand{\beq}{\begin{equation}}
\newcommand{\eeq}{\end{equation}}
\newcommand{\bea}{\begin{eqnarray}}
\newcommand{\eea}{\end{eqnarray}}
\newcommand{\bal}{\begin{align}}
\newcommand{\eal}{\end{align}}

\begin{document}

\title{A Model for the Generation and Transmission of Variations in Evolution}

\author{Olivier Rivoire} 

\address{CNRS/UJF-Grenoble 1, Laboratoire Interdisciplinaire de Physique (LIPhy), UMR 5588, Grenoble, France}

\author{Stanislas Leibler} 

\address{Laboratory of Living Matter, The Rockefeller University, New York, NY, USA}

\address{The Simons Center for Systems Biology, Institute for Advanced Study, Princeton, NJ, USA}

\begin{abstract}
The inheritance of characteristics induced by the environment has often been opposed to the theory of evolution by natural selection. Yet, while evolution by natural selection requires new heritable traits to be produced and transmitted, it does not prescribe, per se, the mechanisms by which this is operated. The mechanisms of inheritance are not, however, unconstrained, since they are themselves subject to natural selection. We introduce a general, analytically solvable mathematical model to compare the adaptive value of different schemes of inheritance. Our model allows for variations to be inherited, randomly produced, or environmentally induced, and, irrespectively, to be either transmitted or not during reproduction. The adaptation of the different schemes for processing variations is quantified for a range of fluctuating environments, following an approach that links quantitative genetics with stochastic control theory.
\end{abstract}

\maketitle

\section{Introduction}

Three principles underlie the explanation of adaptations by natural selection: (i) individuals in a population have varied characteristics;  (ii) their reproductive success correlates with these characteristics; (iii) the characteristics are inherited. The last principle, of inheritance, has always been the most contentious. At the time of Darwin and Wallace, its mechanisms were unknown, and fundamental questions, such as the role of the environment in the production of new, adaptive traits, were unsettled. Adaptation by natural selection does not, indeed, require any causal relation between the environment and newly generated traits, but neither does it exclude it; Darwin, for instance, included as potential sources of variations the direct and indirect effects of the environment, as well as the use and disuse of organs, in line with ideas previously propounded by Lamarck\footnote{The last paragraph of {\it The Origin of Species}~\cite{Darwin59} indicates "Variability from the indirect and direct action of the external conditions of life, and from use and disuse." Darwin's speculations on inheritance, formulated in his hypothesis of pangenesis, were also Lamarckian~\cite{Darwin68}.}.\\
 
Prominent followers of Darwin, however, came to exclude the possibility of inheritance of acquired characteristics. This viewpoint was notably formulated by Weismann in his theory of continuity of the germ-plasm~\cite{Weismann93}. Experiments of amputations, which showed no incidence on the progeny, supported it. At the end of the nineteen century, it had became a central tenet of "neo-Darwinism"~\cite{Romanes95}. Half a century later, the "Modern Synthesis", which produced a synthesis between evolution theory and Mendel's laws of inheritance~\cite{Huxley42}, reached the same conclusion: it promoted a clear distinction between genotypes, inherited but only subject to random variations, and phenotypes, affected by the environment but not directly transmitted. These conclusions were based on studies in multicellular organisms, but subsequent experiments with microorganisms, which found that adaptive variations can precede changes of environmental conditions~\cite{Luria:1943tx}, further reinforced the conviction that biological evolution is mainly fueled by random variations. At a molecular level, finally, once prevalent instructional theories of enzymatic adaptation or antibody formation also came to be discarded in the 1950s and 1960s~\cite{LEDERBERG:1959vq,Monod:1966ta}. At this time, the successes of molecular biology in unraveling the mechanisms of heredity elevated a molecular refutation of Lamarckism, the unidirectional flow of information from DNA to proteins, as its "central dogma"~\cite{Crick:1970wb}.\\

Concurrent views, emphasizing the role of environmentally-induced variations in evolution, have had several insightful proponents~\cite{Baldwin:1896tt,Morgan:1896vr, Osborn:1896vd,Waddington:1942wy}, but were also endorsed by dubious yet influential supporters~\cite{Soyfer94}. Examples of inherited acquired characteristics have however been long known, from the transmission of culture in humans to the uptake of extracellular DNA by bacteria. Yet, only recently have we gained a fuller recognition of the diversity of mechanisms for generating and transmitting variations~\cite{Bonduriansky:2012ka}. In addition to the well-recognized roles of mutations and recombinations of chromosomal DNA, a non-exhaustive list would include the transmission of acquired chromatin marks such as DNA methylation, the transmission of small interfering RNAs, the transmission of conformational states of molecules such as  prions, or, at the cellular level, the transmission of self-sustaining states of gene regulation, and, at the organismal level, so-called parental effects~\cite{Jablonka05}.\\

Inheritance, long treated as an autonomous and universal mechanism to be experimentally characterized and then integrated to evolutionary theory, thus appears to consist of multiple and parallel systems whose origins and implications are to be explained within an evolutionary framework. The problem of a synthesis of inheritance {\it with} evolution is thus now doubled by the problem of the synthesis of inheritance {\it by} evolution. With this problem in view, we propose here a mathematical model where the adaptive values of different schemes for generating and transmitting variations can be compared. The model treats inheritance as a trait on which selection can act, although not in a direct way: systems of inheritance indeed pertain to the transmission of traits between individuals, and estimating their adaptive value therefore requires analyzing the dynamics of a population over several generations\footnote{In this sense, the adaptation of a mode of inheritance is necessarily of "second order" - a form of "evolvability".}.\\

Our model thus quantifies the adaptation of different modes of inheritance by considering the long-term growth rate of populations. The model is both general and analytically solvable. The generality relies on an abstraction from physical implementations, along the example of Shannon's model of communication~\cite{Shannon:1948wk}. The model thus defines the {\it genotype} as what is transmitted between successive generations, and the {\it phenotype} as what determines the survival and reproduction of an individual, with no reference to their material support\footnote{The genotype, defined as what is transmitted between generations, is strictly speaking not a property of an individual, which, formally, only forms a link between genotypes, the genotype that is inherited from the parent(s) and the genotype(s) that is transmitted to the progeny.}. As a consequence, our distinction between genotypic and phenotypic variations is not equivalent to the often made distinction between genetic and epigenetic inheritance; any transmitted character, whether DNA encoded or not, will belong, from the standpoint of our model, to the genotype. \\

In general, few things are excluded in biology if they are not physically impossible - but some have been proposed to be, for instance the absence of reverse flow of information from phenotype to genotype. Regardless of the question of whether this prohibition is indeed universally true or not, it is interesting to consider whether any such prohibition could logically result from natural selection. More generally, under what conditions various mechanisms for generating and transmitting variations may be favored or suppressed by natural selection itself? We illustrate the versatility of our model by examining this question in the context of three biological phenomena that are often considered to be either irrelevant to evolution, or absent because "forbidden":\\

$(i)$ Non-inherited variations, sometimes also referred to as phenotypic "noise", and commonly thought to have no evolutionary implications\footnote{For instance, in the first chapter of {\it The Origin of Species}~\cite{Darwin59}, Darwin states: "Any variation which is not inherited is unimportant for us."}. Considering that new variations may generally be introduced at the genotypic and/or at the phenotypic level, and may thus be transmitted to future generations either totally, in part or not at all, what scheme is most conductive to adaptation? Does natural selection generically favor "developmental canalization", i.e., a reduction of phenotypic differences between individuals inheriting a common genotype?  Our model will highlight how the answer depends on the statistical structure of the environment, and how non-transmitted variations may under some conditions be more beneficial than transmitted variations.\\

$(ii)$ The absence of reverse flow of information from phenotype to genotype, advocated by Weismann, but now challenged even in the species where it is best established~\cite{Sabour:2012hx}. Given that isolating the transmitted genotype from the phenotype may involve dedicated mechanisms, can we characterize the conditions under which natural selection favors their presence?\\

$(iii)$ The non-directed nature of new adaptive variations, associated with the refutation of any "Lamarckian" mechanism. Nothing in principle prevents the environment from inducing the generation of new traits, either at the phenotypic level, or at the genotypic level: the first effect, known as "plasticity", has long been recognized~\cite{WestEberhard03}, and the second one, long thought to be forbidden, is also observed~\cite{Jablonka:98}. Are there nevertheless conditions under which direct integration of information into the transmitted genotype is logically excluded as a consequence of natural selection?\\

Despite the fundamental nature of these questions, no previous formal model exists, to our knowledge, that addresses them in a common and analytically tractable framework. Our model, however, is not without precedents: it is in line with traditional models of quantitative genetics~\cite{Fisher30}, and relates to models of stochastic control in engineering~\cite{Maybeck:1979vh}. Processing unreliable informations from the past and present to confront an uncertain future, which may be seen as the fundamental "function" of systems of inheritance, is indeed at its core a question of control. We thus proposed previously that evolution in fluctuating environments could be viewed as a problem of stochastic control~\cite{Rivoire:2011ue}; this view is supported here by a formal analogy between our model and a basic algorithm in stochastic control theory, the Kalman filter~\cite{Kalman:1960tn}.

\section{Model}

We provide in this section a general presentation of the model and derive its solution in a simple case. The general solution follows the same principles and its details are included as supplementary information.

\subsection{Definition}

The model considers a population of asexually reproducing individuals where each genotype is characterized by an "attribute" $\g$. The genetic or epigenetic nature of this attribute is irrelevant\footnote{It might be more natural to simply call this quantity a "trait", except that this term is often used as an abbreviation for "phenotypic trait", and we want to emphasize that $\g$ is the transmitted information, i.e., the genotype. In the parlance of Weismann, $\g$ would be termed the "germ-plasm", and, in the parlance of population genetics, the "breeding value".}: we are only concerned with the origins of the transmitted information, either inherited, randomly produced, or environmentally induced, irrespective of its material support. At each time step, corresponding to a generation, each individual with attribute $\g$ reproduces and is replaced by $\xi$ offsprings sharing as common\footnote{The model could be generalized to produce offsprings with different attributes, but we purposely ignore this unessential complication; anecdotally, we may also note that "polyembryony", the reproduction into genetically identical siblings, does occur systematically in some species such as armadillos~\cite{Loughry:1998vp}.} attribute $\g'$ (with possibly $\xi=0$, in which case we conventionally define $\g'=\g$). The individuals are non-interacting and the generations non-overlaping. The values of $\xi$ and $\g'$ can depend on $\g$ and on the current environmental state, which fluctuates independently of the population and is characterized by a variable $x_t$. This dependency can be stochastic, and is generally given by a stochastic kernel $A(\xi,\g'|\g,x_t)$ with the following properties: $A(\xi,\g'|\g,x_t) \ge 0$ and $\sum_\xi \int \ud \g'\ A(\xi,\g'|\g,x_t)=1$ for all $\g,x_t$.

\subsection{Population dynamics}\label{sec:whyL}

For a given series of environmental states $(x_1,\dots, x_T)$, we define $n_t(\g)$ as the expected probability density function of the attribute $\g$ in the population at time (generation) $t$, normalized to $\int \ud \g\ n_t(\g)=1$ at any $t$. It satisfies the recursive equation\footnote{More formally, this equation is the recursion for the first moment  of the branching process.}
\beq\label{eq:meanfield}
n_{t+1}(\g')=W_t^{-1}\int\ud \g\ \sum_{\xi}\ \xi\ A(\xi,\g'|\g,x_t)\ n_t(\g),
\eeq
where $W_t$ is a normalization ensuring $\int \ud \g'\ n_{t+1}(\g')=1$,
\beq
W_t=\int\ud\g'\int\ud \g\ \sum_{\xi}\ \xi\  A(\xi,\g'|\g,x_t)\ n_t(\g).
\eeq
$W_t$ represents the factor by which the population increases, on average, between generations $t$ and $t+1$.\\

Starting from a large number $N_0$ of individuals at time $t=0$, the expected total number $N_T$ of individuals at time $T$ is thus
\beq
N_T=\prod_{t=1}^TW_t\ N_0.
\eeq
Over $T$ generations, this results in the growth rate
\beq
\Lambda\equiv\frac{1}{T}\ln\frac{N_T}{N_0}=\frac{1}{T}\sum_{t=1}^T\ln W_t.
\eeq
We are interested here in the long-term limit $T\to\infty$, under the assumption that the population does not go extinct\footnote{We assume that the branching process is supercritical and ignore the fluctuations associated with small populations, which is justified in the large $t$ limit when the population is exponentially growing.}. This limit is mathematically well defined when the environment follows an ergodic process, in which case we have
\beq
\Lambda=\lim_{t\to\infty}\E[\ln W_t],
\eeq
where $\E$ indicates an expectation with respect to the environmental fluctuations~\cite{Tuljapurkar90} ($\Lambda$ is also known as the quenched Lyapunov exponent for the underlying branching process in a random environment).\\

The long-term growth rate $\Lambda$ is a "group-level" property, attached to the population as a whole rather than to any particular individual. It is relevant for the comparison of different schemes of inheritance because of the following property~\cite{Rivoire:2011ue}: given two populations, characterized by kernels $A_1$ and $A_2$, and given an ergodic environmental process, if $\Lambda(A_1)>\Lambda(A_2)>0$ then, almost surely, $\lim_{t\to\infty}\ln N_t^2/N_t^1=0$, where $N_t^1$ and $N_t^2$ represent the respective sizes of the two populations at time $t$. In other words, $\Lambda$ predicts the long-term outcome of a competition between populations characterized by different kernels $A$. This argument assumes an exponentially growing population, but its conclusions are equally valid in presence of a constraint on the total population size, in which case the population with smallest growth rate almost surely becomes extinct.

\subsection{Long-term growth rate}\label{sec:simple}

\begin{figure}[t]
\begin{center}
\includegraphics[width=\linewidth]{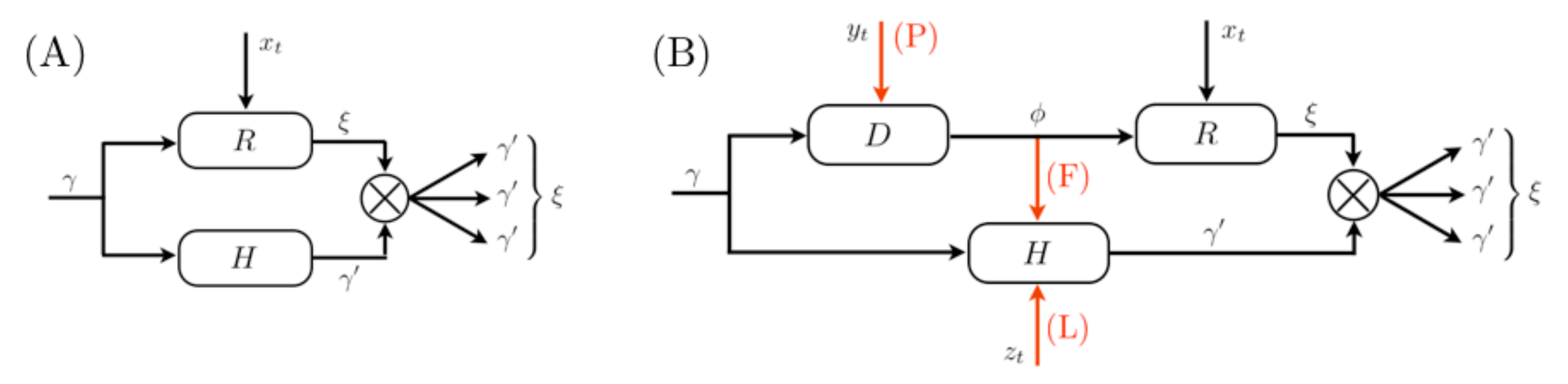}
\caption{{\bf (A)} Simple model -- In this model, no distinction is made between inherited genotype and phenotype: both are characterized by a single "attribute" $\g$. The number $\xi$ of offsprings and the attribute $\g'$ that they transmit are independently specified by two stochastic kernels, the "reproduction kernel" $R(\xi|\g,x_t)$ and the "heritability kernel" $H(\g'|\g)$, where $\g$ represents the attribute inherited by the individual and $x_t$ the current state of the environment. {\bf (B)} General model -- This model includes explicitly a phenotype $\phi$ derived from the inherited genotype $\g$ through a stochastic kernel $D(\phi|\g,y_t)$ representing "development". This kernel, and the kernels $R(\xi|\g,x_t)$ and $H(\g'|\g,\phi,z_t)$, can depend on external, environmental factors, $x_t$, $y_t$, $z_t$. The three red arrows represent respectively developmental plasticity ({P}), phenotype-to-genotype feedback (F) and direct Lamarckian effects (L).\label{fig:scheme0}}
\end{center}
\end{figure}

Different systems of inheritance are represented in the model by different kernels $A$. In the simplest case, schematically represented in Fig.~\ref{fig:scheme0}A, the environment affects reproduction but not transmission, and the kernel $A$ is factorized into the product of a reproduction kernel $R$ and an heredity kernel $H$,
\beq\label{eq:simplest}
A(\xi,\g'|\g,x_t)=R(\xi|\g,x_t)H(\g'|\g).
\eeq
From now on, for simplicity, we also assume a single continuous attribute $\g\in\mathbb{R}$, although the model remains analytically solvable in the multi-dimensional case.\\

We take for the heredity kernel
\beq\label{eq:H}
H(\g'|\g)=G_{\sq_H}(\g'-\g),
\eeq
where $G_{\sq}(x)$ denotes a generic Gaussian function with variance $\sq$,
\beq
G_{\sq}(x)\equiv\frac{1}{(2\pi\sq)^{1/2}}\exp\left({-\frac{x^2}{2\sq}}\right).
\eeq
 Equivalently, we may write the relation between the transmitted attribute $\g'$ and the inherited attribute $\g$ as
\beq
\g'=\g+\eta,\quad\textrm{with}\quad \eta\sim\N(0,\s_H^2),
\eeq
where $\N(0,\sq_H)$ denotes a normal distribution with variance $\sq_H$. This corresponds to a standard assumption of additivity in population genetics (extended below to include a possible reversion towards a mean, i.e., $\g'=\l\g+\eta$ with $\l<1$).\\

To compute $n_t(\g)$, as in Eq.~\eqref{eq:meanfield}, we only need to specify the first moment of the selection kernel $R(\xi|\g,x_t)$. We assume here that the expected number of offsprings of an individual with attribute $\g$ in environment $x_t$ is
\beq\label{eq:R}
\langle\xi\rangle_{\g,x_t}\equiv\sum_{\xi}\xi\ R(\xi|\g,x_t)=\exp\left(r_{\rm max}-\frac{(\g-x_t)^2}{2\sq_S}\right).
\eeq
Here, the difference $\g-x_t$ captures the "fitness" of an individual with attribute $\g$ to the current criterion of selection $x_t$. Since in the simplest version of the model, depicted in Fig.~\ref{fig:scheme0}A, there is no distinction between inherited genotype and phenotype, the attribute $\g$ can be thought as a phenotypic trait and the "state of the environment" $x_t$ is thus being defined relative to the population, as the value of the trait that is optimal in this environment (for instance, in a classical simplistic picture of adaptation, $x_t$ would be associated with the height of acacias and $\g$ with the length of the giraffe neck). The variance $\sq_S$ describes the selectivity of the environment; $\sq_S=0$ means that only one phenotype can survive at any given time, while $\sq_S$ large means that many different phenotypes can survive. Finally, $r_{\rm max}$ is the maximal reproductive rate per generation for the species; in particular, $r_{\rm max}>0$ is a necessary condition for the population not to go extinct\footnote{$r_{\rm max}$ affects the growth rate $\Lambda$ only trough an additive constant, i.e., $\Lambda(r_{\rm max})=\Lambda(r_{\rm max}=0)+r_{\rm max}$, and therefore plays no role when comparing different inheritance schemes.}.\\

Starting at $t=0$ from a large population with an normally distributed attribute, $n_0(\g_0)=G_{\sq_0}(\g-m_0)$, the distribution of $\g$ remains normally distributed at all times, i.e., $n_t(\g)=G_{\sq_t}(\g-m_t)$ for all $t$; more generally, starting from any distribution $n_0(\g_0)$, the distribution of the trait in the population will converge to a Gaussian. Assuming that it does not go extinct, the long-term evolution of the population can thus be described in terms of just two parameters, the mean $m_t\equiv\langle\g\rangle_t$ and variance $\sq_t\equiv\langle(\g-m_t)^2\rangle_t$ of $\g$ at time $t$.\\

From Eq.~\eqref{eq:meanfield}, it follows that these two parameters satisfy
\bea
m_{t+1}&=&m_t+h_t^2(x_t-m_t),\label{eq:recm}\label{eq:rec1}\\
\sq_{t+1}&=&h_t^2+\sq_H,\label{eq:rec2}
\eea
where the so-called heritability $h_t^2$, defined by
\beq
h_t^2=\frac{\sq_t}{\sq_S+\sq_t},
\eeq
satisfies the recursion
\beq\label{eq:rech}
h_{t+1}^2=1-\frac{1}{1+(\sq_H+h^2_t)/\sq_S}.
\eeq
Eq.~\eqref{eq:meanfield} also yields in terms of these variables $W_t$, the factor by which the population size increases between times $t$ and $t+1$, 
\beq\label{eq:lnW}
\ln W_t=\frac{1}{2}\ln\left(1-h_t^2\right)-\frac{1}{2}\left(1-h_t^2\right)\frac{(m_t-x_t)^2}{\sq_S}.
\eeq

\subsection{Stochastically fluctuating environments}

Up to here, the equations are valid for any kind of environmental process. If we now assume an ergodic environment, we obtain, by taking the $t\to\infty$ limit,  a formal expression for the long-term growth rate,
\beq\label{eq:expr}
\Lambda=\frac{1}{2}\ln\left(1-h_\infty^2\right)-\frac{1}{2}\left(1-h_\infty^2\right)\lim_{t\to\infty}\frac{\E[(m_t-x_t)^2]}{\sq_S},
\eeq
where $h_\infty$ represents the fixed point of $h_t$ in Eq.~\eqref{eq:rech}. $\Lambda$ is the sum of two terms: the first can be interpreted as the "genetic load" due to stabilizing selection and the second as the "evolutionary load" due to the lag between the mean phenotype $m_t$ and the optimal phenotype $x_t$. 
The dynamics of $m_t$ has generally no fixed point (unless the environment is constant), but, if the environment is ergodic, it has a stationary distribution and hence $\E[(m_t-x_t)^2]$ has a limit for $t\to\infty$.\\

$\Lambda$ may be explicitly computed for several stationary processes. Two aspects of the environment are particularly relevant when studying the adaptive value of mechanisms of heredity: the amplitude $\sq_E$ of the environmental fluctuations, and the scale $\tau_E$ of their temporal correlations. As a simple dynamical process that encapsulates these two elements we consider
\beq
 x_t=a x_{t-1}+b_t,\qquad\textrm{with}\qquad b_t\sim\N(0,\sq_X).
 \eeq
It corresponds to $\{x_t\}_t$ being generated by a stationary Gaussian Markov process with transition kernel $P(x_t|x_{t-1})=G_{\sq_X}(x_t-ax_{t-1})$.
The parameter $\sq_X$ controls the degree of stochasticity and the parameter $a$ the degree of correlation between successive environments; assuming $a\leq 0<1$, the process followed by $x_t$ has a stationary distribution, namely $\N(0,\sq_X(1-a^2)^{-1})$. The amplitude $\sq_E$ and relaxation time $\tau_E$ of this process are thus defined by
\beq
\sq_E\equiv\frac{\sq_X}{1-a^2},\qquad\tau_E\equiv -\frac{1}{\ln a},
\eeq
such that\footnote{From the definition $x_{t+1}=ax_t+b_t$ with $\E[b_t]=0$ and $\E[b_t^2]=\sq_X$, and the assumption that $x_t$ has a stationary distribution, we have indeed $\sq_E\equiv\E[x^2_t]=\E[x_{t+1}^2]=a^2\E[x^2_t]+\E[b_t^2]=a^2\sq_E+\sq_X$, and therefore $\sq_E=\sq_X/(1-a^2)$. And from $\E[x_{t+t'}x_{t'}]=a^t\E[x_{t'}^2]$, we obtain $\E[x_{t+t'}x_{t'}]=\sq_Ee^{-t/\tau_E}$ with $\tau_E=-(\ln a)^{-1}$.}
\beq
\E[x_{t+t'}x_{t'}]=\sq_E\ e^{-t/\tau_E}.
\eeq
This stochastic process is known as an autoregressive AR(1) model in signal processing and corresponds in physics to a discrete-time Ornstein-Uhlenbeck process.\\

With this choice for the environmental temporal dynamics, we can show that $m_t$ itself is normally distributed and we can compute $\lim_{t\to\infty}\E[(m_t-x_t)^2]$ (see Suppl. Info.). The calculation leads to a long-term growth rate $\Lambda$ of the form
\beq
\L=r_{\rm max}+\L_0\left(a,\frac{\sq_H}{\sq_S},\frac{\sq_E}{\sq_S}\right).
\eeq
Without loss of generality, we can therefore assume that $\sq_S=1$. Under this assumption, the expression for $\L_0$ is given by
\beq\label{eq:Lsimple}
\Lambda_0(a,\sq_H,\sq_E)=\frac{1}{2}\ln\alpha-\frac{(1-a)\alpha}{(1+\alpha)(1-a\alpha)}\sq_E,\qquad\textrm{with}\qquad \alpha=\frac{2}{2+\sq_H+\left(\sq_H(\sq_H+4)\right)^{1/2}}
\eeq
where the relation between $\alpha$ and $\sq_H$ can also be inverted to give $\sq_H=(1-\a)^2/\a$.\\

Not considering the additive parameter $r_{\rm max}$, only three parameters, $a$, $\sq_E$, $\sq_H$, are needed to characterize this simple model. The first two parameters pertain to the environmental process, with $a$ representing temporal correlations ($a=e^{-1/\tau_E}$) and $\sq_E$ its stationary variance (in units of $\sq_S$), while $\sq_H$ (in the same units) describes the stochasticity of inheritance (akin to a mutation rate). We take the view that the parameter $\sq_H$ is evolving on a slow time scale, relative to the time scale $\tau_E$ of the environment, so that the outcome of a competition between populations with different $\sq_H$ is determined by $\Lambda(\sq_H)$ (adiabatic limit). How $\Lambda$ depends on the different parameters is, however, not evident from the formulae. We examine this issue in Sec.~\ref{sec:selected} after introducing a generalization of the model that makes it suitable to addressing the questions raised in the introduction (the model of this section is treated in Suppl. Info., see Fig.~S1).

\subsection{Generalization}\label{sec:general}

The previous formulae solve the simple model schematically represented in Fig.~\ref{fig:scheme0}A. A generalization of this model, depicted in Fig.~\ref{fig:scheme0}B,  explicitly distinguishes between the inherited genotype $\g$ and the phenotype $\phi$ of an individual. The phenotype arises from the inherited genotype through a process of "development", represented by a stochastic kernel $D$,  which can  show some dependence on the external environment, i.e., developmental plasticity (arrow P in Fig.~\ref{fig:scheme0}B). The generalized model also incorporates the possibility of  acquiring information from the environment and directly modifying the heritability kernel $H$ (arrow L). The latter can also be influenced internally by a feedback from the phenotype to the transmitted genotype (arrow F). This more general model, several limits of which we analyze in the following sections, can also be solved analytically.\\

More precisely, the general model is characterized by the following equations:
\bea
\g'&=&\lambda\g+\kappa z_t+\omega\phi+\nu_H,\qquad\qquad\nu_H\sim\N(0,\sq_H),\\
\phi&=&\theta\g+\rho y_t+\nu_D,\qquad\qquad\qquad\ \ \nu_D\sim\N(0,\sq_D),\\
\langle\xi\rangle_{\g,x_t}&=&\exp[r_{\rm max}-(\phi-x_t)^2/(2\sq_S)].
\eea
The first equation specifies how the transmitted genotype $\g'$ depends on the inherited genotype $\g$, on some external information $z_t$ coming from the environment and on the current phenotype $\phi$. The second equation defines how this current phenotype depends on the inherited genotype $\g$ and on some possibly available external information $y_t$. The third equation, finally, gives the expected number of offsprings for individuals with phenotype $\phi$ in environment $x_t$.\\

The description of the model is completed by the equations governing the dynamics of the environment. As before, it is supposed to be fixed (quenched) independently of the dynamics of the population:
\bea
x_t&=&ax_{t-1}+b_t,\quad\ \  b_t\sim\N(0,\sq_X),\\
y_t&=&x_t+b'_t,\quad\qquad\ b'_t\sim\N(0,\sq_{Ip}),\\
z_t&=&x_t+b''_t,\quad\qquad b''_t\sim\N(0,\sq_{I\ell}).
\eea
The signals $y_t$ and $z_t$ are thus assumed to derive from $x_t$ via additive white Gaussian noise channels, one of the simplest models of signal transmission.\\

Although involving more parameters, the derivation of the solution for this general model follows the same principles as for the simple model presented previously\footnote{The solution of this simpler model can be recovered by taking $\l=1$, $\theta=1$, $\sq_D=0$, $\k=0$, $\omega=0$, $\rho=0$.}. Here again, we can assume that all variances are expressed in units of $\sq_S$, or, equivalently, assume that $\sq_S=1$. We thus obtain
\beq\label{eq:general}
\begin{split}
\Lambda= &\ r_{\rm max}+\frac{1}{2}\ln\left(\frac{\alpha}{\tilde\lambda(\sq_D+1)}\right)-\frac{1}{2}\frac{\alpha}{\tilde\l(\sq_D+1)}\\
&\times\left[(1-\rho^2)\frac{(1+a\alpha)(1+(\tilde\l+\tilde\k)^2)-2(a+\alpha)(\tilde\l+\tilde\k)}{(1-a\alpha)(1-\alpha^2)}\sq_E+\rho^2\frac{1+(\tilde\l+\tilde\omega)^2-2\alpha(\tilde\l+\tilde\omega)}{1-\alpha^2}\sq_{Ip}+\kappa^2\theta^2\sq_{I\ell}\right],
\end{split}
\eeq
where
\beq
\begin{split}
\a&=\frac{2\tilde\lambda}{1+\tilde\lambda^2+\tilde\sigma_H^2+\left((1-\tilde\lambda^2-\tilde\sigma_H^2)^2+4\tilde\sigma_H^2\right)^{1/2}},\qquad\tilde\sigma_H^2=\left(\sq_H+\frac{\omega^2\sq_D}{\sq_D+1}\right)\frac{\theta^{2}}{\sq_D+1},\\
\tilde\l&=\l+\frac{\theta\omega}{\sq_D+1},\qquad\tilde\k=\left(\frac{\sq_D+\rho}{\sq_D+1}\omega+\kappa\right)\frac{\theta}{1-\rho},\qquad\tilde\omega=\frac{(\sq_D-1)\theta\omega}{(\sq_D+1)(1-\rho)}.
\end{split}
\eeq
We verified the validity of this analytic expression by comparisons with numerical simulations of the dynamics of the population for various values of the parameters. We use this result to analyze three particular variants of the model, which address the three specific questions raised in the introduction. In each case, we characterize the most adaptive scheme for generating and transmitting variations by considering the value of the parameters that optimize $\Lambda$ - the expected outcome of an evolutionary dynamics where these parameters evolve on a time scale longer than the characteristics time scale $\tau_E$ of the environment\footnote{It can be shown that $\Lambda$ is concave with respect to its parameters~\cite{Cohen:1980}, implying an absence of local maxima into which the evolutionary dynamics could otherwise be trapped.}.

\section{Three selected questions}\label{sec:selected}

\subsection{Where to introduce variation?}\label{sec:wherevar}

The introduction of new variations is a requirement for sustained evolution: a population into which no new variations are introduced will eventually become monomorphic even in absence of selection, simply as a consequence of "random drift". On the other hand, if the characters of the individuals are so new as to be uncorrelated with those of their parents, inheritance is negated, and adaptation by natural selection impossible. Therefore, an appropriate "degree" of new variations must lie between these two extremes. This logical conclusion raises a first class of questions: What is this intermediate degree of variations? What sets its value? And if an optimal degree exists, can it be selected for? Several biological observations hint at a positive answer to this last point: a complex molecular machinery has for instance evolved to ensure a faithful replication of DNA, and thus to "set" its mutation rate; moreover, not all genes are treated equally: in bacteria, for example, the positioning of genes on the leading or lagging strand of the chromosome, where they are subject to different mutation rates, is correlated to the nature of the selective pressure that they experience~\cite{Paul:2014em}.\\

A second class of questions follows from noticing that new variations may not only vary in "degree" but also in "nature". In particular, new variations may be phenotypic, thus affecting survival and reproduction but not being transmitted to the next generation, or/and genotypic, thus being transmitted to the next generation but not directly affecting survival and reproduction. How different are these two (non-exclusive) types of variation from an evolutionary standpoint? Is one type of variations more advantageous than the other? Or, does each have its own optimal "degree"? Here again, several observations support the biological relevance of these questions. Both types of variations are indeed  found simultaneously in every living organism. An example of universally shared genotypic variations is provided by DNA mutations, while the so-called non-genetic individuality of bacteria~\cite{Spudich:1976wv} is an example of  phenotypic variations. The latter, while often assimilated to "noise", may in fact confer a selective advantage to the organisms~\cite{Eldar:2010kk}.\\

\begin{figure}[t]
\begin{center}
\includegraphics[width=.5\linewidth]{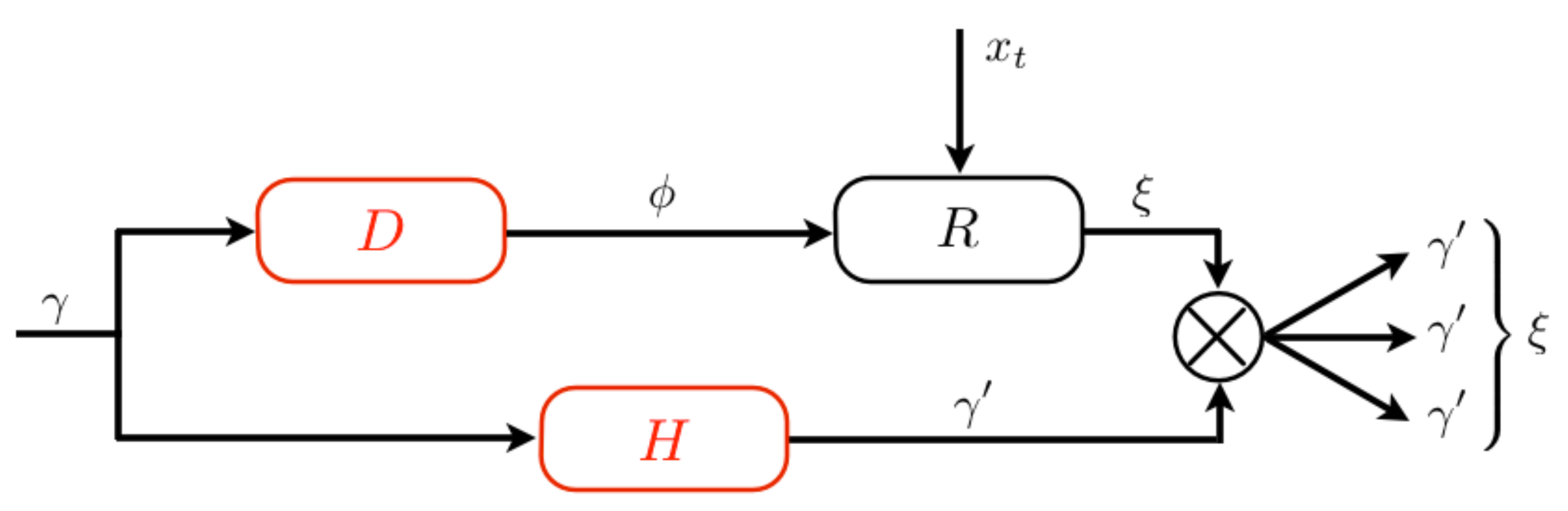}
\caption{Where to introduce variation? This question is addressed within a model where the two sources of noise, the developmental kernel $D(\phi|\g)$ and heredity kernel $H(\g'|\g)$, are jointly optimized to yield the largest long-term growth rate. The optimization is performed over the two parameters $\sq_D$ and $\sq_H$ which define $D(\phi|\g)$ and $H(\g'|\g)$ by the relations $\phi=\g+\nu_D$ and $\g'=\g+\nu_H$, with $\nu_D\sim\N(0,\sq_D)$ and $\nu_H\sim\N(0,\sq_H)$.\label{fig:scheme1}}
\end{center}
\end{figure}

 Addressing questions about the "degree" and "nature" of new variations requires a model that is both quantitative and rich enough to allow for non-trivial selective pressures. Our model meets these criteria; in fact, it is even sufficient to consider its simplified version, in which all three mechanisms depicted by red arrows in Fig.~\ref{fig:scheme0}B, namely plasticity, feedback and Lamarckian effects, are absent. This limiting case, represented in Fig.~\ref{fig:scheme1}, is described by the equations
\begin{align}
\g'&=\g+\nu_H,\qquad\nu_H\sim\N(0,\sq_H),\nonumber\\
\phi&=\g+\nu_D,\qquad\nu_D\sim\N(0,\sq_D).\label{eq:model1}
\end{align}
In this particular instance of the model\footnote{Obtained from the general model by taking $\l=1$, $\theta=1$, $\k=0$, $\omega=0$, $\rho=0$ (we may consider more generally arbitrary values of $\l$ and $\theta$ at the expense of introducing more parameters).}, new variations are generated independently of the state of the environment and are introduced at two levels: at the genotypic level, through the random variable $\nu_H$ with variance $\sq_H$, and at the phenotypic level, through the random variable $\nu_D$ with variance $\sq_D$. The two possible types of new variations are thus tunable at various degrees. By studying how the long-term growth $\Lambda$ may be optimized with respect to $\sq_H$ and $\sq_D$, we can thus analyze quantitatively the adaptive value of these two sources of variation as a function of the characteristics of the environment, its temporal correlation $a$ and its stochasticity $\sq_E$.\\

\begin{figure}[t]
\begin{center}
\includegraphics[width=.9\linewidth]{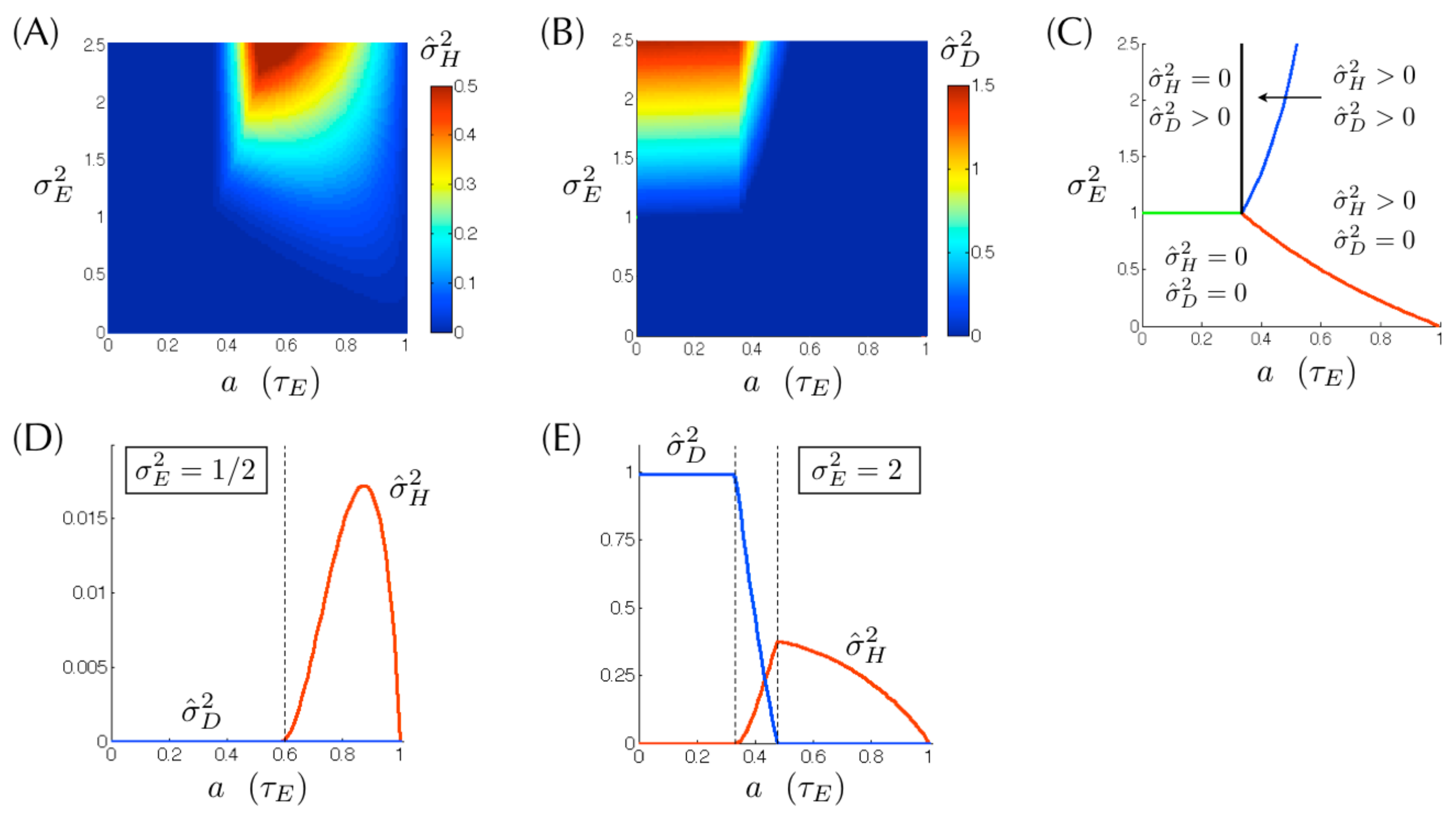}
\caption{Where to introduce variation? -- {\bf (A)} For the model described by Eqs.~\eqref{eq:model1} and depicted in Fig.~\ref{fig:scheme1}, optimal values of the degree $\sigma_H^2$ of genotypic variations as a function of the environmental variables $(a,\sigma_E^2)$ when optimizing jointly over $(\sigma_H^2,\sigma_D^2)$. $\sigma_E^2$ represents the amplitude of the environmental fluctuations and $a$, via $\tau_E=-(\ln a)^{-1}$, the characteristic time scale of their relaxation. {\bf (B)} Optimal values of the degree $\sigma_D^2$ of phenotypic variations as a function of $(a,\sigma_E^2)$ when optimizing jointly over $(\sigma_H^2,\sigma_D^2)$. {\bf ({C})} "Phase diagram" for the optimal values of $(\sigma_H^2,\sigma_D^2)$ as a function of $(a,\sigma_E^2)$. {\bf (D)} Optimal values of $(\sigma_H^2,\sigma_D^2)$ as a function of $a$ for $\sigma_E^2=1/2$. {\bf (E)} Optimal values of $(\sigma_H^2,\sigma_D^2)$ as a function of $a$ for $\sigma_E^2=2$.}\label{fig:FigQ1}
\end{center}
\end{figure}

Specifically, we analyze here the values ($\hat\sigma_H^2$, $\hat\sigma_D^2$) of the variables $\sigma_H^2 $ and $\sigma_D^2$ that optimize the long-term growth $\Lambda$, for given values of $(a,\sigma_E^2)$ (fixing without loss of generality $\sq_S=1$); this corresponds to the expected outcome of a competition between populations characterized by different values of $\sigma_H^2 $ and $\sigma_D^2$, or, equivalently, to the expected outcome of an evolutionary dynamics where $\sigma_H^2 $ and $\sigma_D^2$ are themselves slowly varying. Numerical results, presented in Fig.~\ref{fig:FigQ1}A-B, show that the nature of the most adaptive variations indeed depends on the statistics of the environment. For instance, phenotypic noise ($\sigma_D^2>0$) is preferred over genotypic noise ($\sigma_H^2>0$) for weakly correlated environments (low $a$). The optimization can in fact be performed analytically to derive a phase diagram with distinct phases, defined by the presence or absence of phenotypic or genotypic stochasticity in the optimal solution. The different boundaries, shown in Fig.~\ref{fig:FigQ1}C, are given by
\beq\label{eq:boundaries}
a=1/3\quad {\rm (black)},\qquad\sigma_E^2=\frac{1}{4}\left(\frac{1+a}{1-a}\right)^2\quad {\rm (blue)},\qquad
\sigma_E^2=2\left(\frac{1-a}{1+a}\right)\quad {\rm (red)},\qquad\sigma_E^2=1\quad {\rm (green)},
\eeq
with the central point at $a=1/3$ and $\sq_E=1$. When $\sigma_E^2<1$, as in Fig.~\ref{fig:FigQ1}D, we thus encounter two phases as $a$ is varied; when $\sigma_E^2>1$, as in Fig.~\ref{fig:FigQ1}E, three phases are traversed.\\

A first conclusion from these results is that a population can be adapted to a fluctuating environment (here have optimal $\sq_H$ and $\sq_D$) even though its individuals are not undergoing any change: this is the case for the regions of the phase diagram where $\hat\sigma_H^2=0$, corresponding to a population with a homogeneous and constant genotype. In this sense, adaptation of a population to a fluctuating environment does not require variations in the attributes of the individuals. Yet, an absence of evolution does not necessarily imply an absence of diversity. When $\hat\sigma_H^2=0$ but $\hat\sigma_D^2>0$, the population is phenotypically diverse, although rather than transmitted from one generation to the next, the same diversity is reproduced at each generation. Moreover, a population may also be adapted to a fluctuating environment without showing any diversity: for small environment fluctuations, we find indeed that $\hat\sigma_H^2=0$ and $\hat\sigma_D^2=0$ (see Fig.~\ref{fig:FigQ1}C); in this case, natural selection favors the suppression of any variation\footnote{The diversity of the population, either genotypic or phenotypic, can be more precisely quantified within the model. We thus have 
$$\varsigma_\g\equiv\frac{\E[(\g-m_t)^2]}{\sq_S}=\frac{1-\a}{\a},$$
for the genotypic diversity, and, since $\E[(\phi-m_t)^2]=\E[(\g-m_t)^2]+\E[(\g-\phi)^2]$,
$$\varsigma_\phi\equiv\frac{\E[(\phi-m_t)^2]}{\sq_S}=\frac{1-\a}{\a}+\sigma_D^2.$$
for the phenotypic diversity. Since $\sigma_H^2=0$ corresponds to $\alpha=1$, we verify with these formula that $\varsigma_\g=0$ if and only if $\sigma_H^2=0$, and $\varsigma_\phi=0$ if and only if $\sigma_H^2=0$ and $\sigma_D^2=0$.}.\\

While new variations are beneficial when the fluctuations of the environment are large enough, our model predicts that natural selection should favor their introduction at different levels, depending on the statistical structure of these fluctuations. As indicated in Fig.~\ref{fig:FigQ1}C, phenotypic variations are suppressed ($\hat\sigma_D^2=0$) when the environmental stochasticity is small enough or the environmental correlation large enough: this may be interpreted as a selection for "canalization", i.e., reduction of the phenotypic diversity of genotypically identical individuals, a phenomenon indeed observed in biological organisms~\cite{Waddington:1942wy}. Genotypic variations are suppressed ($\hat\sigma_H^2=0$) on the other hand, when the environment is not strongly correlated ($a$ small). This may be rationalized by noticing that non-trivial inheritance is relevant only when successive generations share correlated selective pressures.\\

Living organisms do not harbor a single trait, but many, each potentially subject to a selective pressure with a different statistical structure. For instance, in bacteria, the strength of selection may be very different between central metabolism and mechanisms of resistance to antibiotics. Our model suggests that this diversity of selective pressures may be responsible for the evolution of the diversity of ways in which new traits are generated and transmitted.  However, our model is obviously extremely schematic and does not account for a number of features that affect the evolution of mechanisms of inheritance. In particular, it does not consider the cost of these mechanisms, which may strongly limit their actual diversity: suppressing any variation by error corrections, checkpoints, canalization, etc. may be prohibitively expensive, and evolving a different system of inheritance for every trait simply impossible (any mechanism for introducing variations in a trait defines a new trait into which variations may be introduced).

\subsection{When to separate phenotype and transmitted genotype?}\label{sec:whengp}

The previous version of the model assumes an independent germ-line, with a transmitted genotype $\g'$ that is not influenced by the phenotype $\phi$. Such a separation between a "germ-plasm" $\g$ and a "soma" $\phi$ is central to the view that new traits are exclusively generated by random mutations in the gametes, independently of any event occurring during the life-time of the individual. This separation, however, cannot be taken for granted, and is in fact absent in many if not most living organisms, including notably plants~\cite{Whitham:1981uu}. Yet, mammals do seem to possess specific mechanisms to enforce a separation; for instance, murine primordial germ cells undergo resetting and erasing of maternal and paternal imprints, genome-wide DNA methylation, extensive histone modifications, and inactive X-chromosome reactivation~\cite{Sabour:2012hx}. As a very first step in trying to understand the origin of such mechanisms, it is instructive to abstract from the many constraints that may limit the evolution of systems of inheritance, and look for the way in which genotypic and phenotypic features should ideally be combined to ensure a maximal growth rate of the population; if a "Weismann's barrier" segregating a germ line from the soma is never found in such conditions, this implies that its origin must reside elsewhere. We follow here this approach by examining within model the selective value of a feedback from phenotype to transmitted genotype.\\

\begin{figure}[t]
\begin{center}
\includegraphics[width=.5\linewidth]{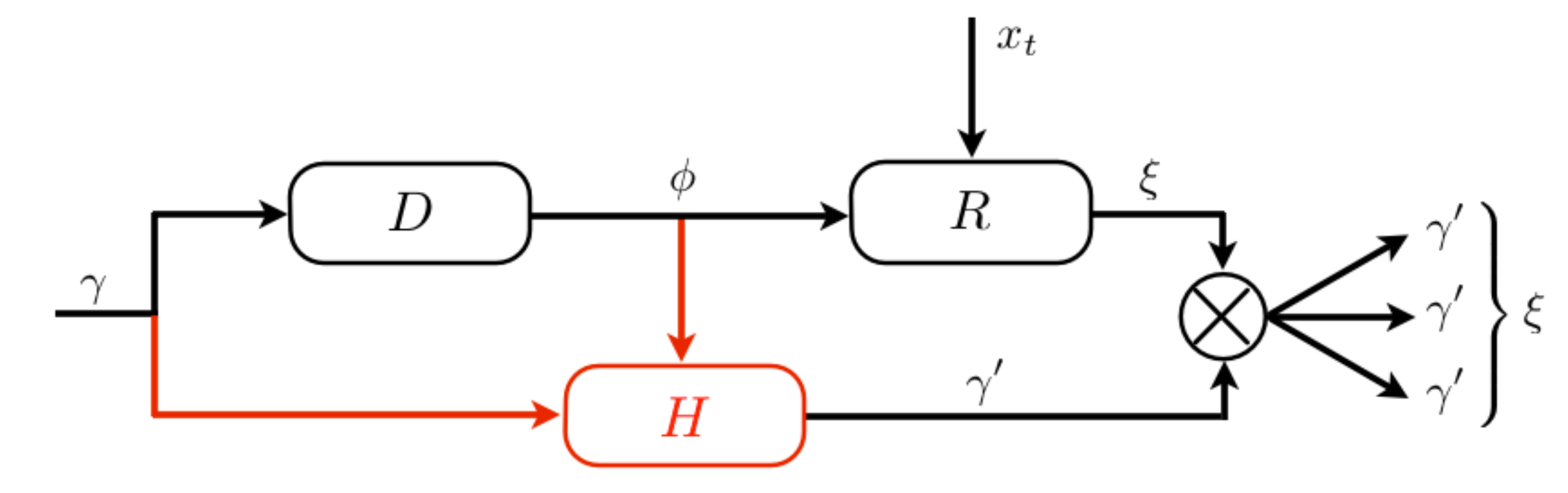}
\caption{When to separate phenotype and transmitted genotype? This question is addressed within a model where the heredity kernel $H(\g'|\g,\phi)$ is optimized. The optimization is made over the three parameters $\l$, $\omega$, $\sq_H$ which define $H(\g'|\g,\phi)$ by the relation $\g'=\l\g+\omega\phi+\nu_H$ with $\nu_H\sim\N(0,\sq_H)$.\label{fig:scheme2}}
\end{center}
\end{figure}

\begin{figure}[t]
\begin{center}
\includegraphics[width=.9\linewidth]{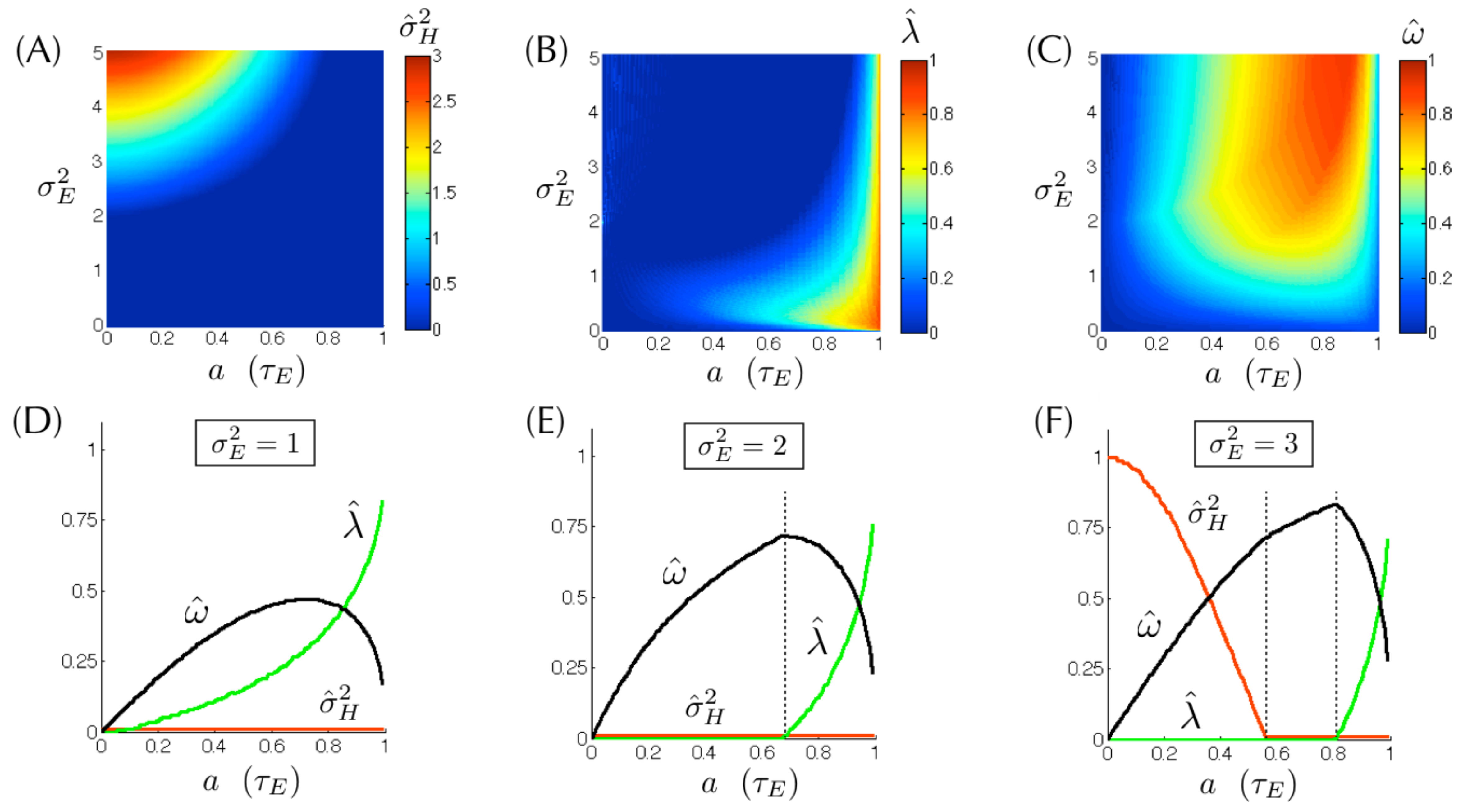}
\caption{When to separate phenotype and transmitted genotype? -- {\bf (A-C)} For the model described by Eqs.~\eqref{eq:model2} and depicted in Fig.~\ref{fig:scheme2}, the values of $(\sigma_H^2,\lambda,\omega)$ that jointly optimize $\Lambda$ are represented as a function of the environmental parameters $(a,\sigma_E^2)$ for a fixed developmental noise $\sigma_D^2=1$. {\bf (D-F)} Same results presented as a function of $a$ for three fixed values of $\sigma_E^2$.}\label{fig:QA}
\end{center}
\end{figure}

Three factors potentially contribute to the genotype $\g'$ transmitted to the offsprings: the genotype $\g$ inherited from the parent, the phenotype $\phi$ of the individual, and random variations $\nu_H$. To analyze their relative adaptive value, we study the model depicted by Fig.~\ref{fig:scheme2}, which allows for different combinations of these three elements:
\begin{align}
\g'&=\lambda\g+\omega\phi+\nu_H,\qquad\nu_H\sim\N(0,\sq_H),\nonumber\\
\phi&=\g+\nu_D,\qquad\qquad\quad\nu_D\sim\N(0,\sq_D).\label{eq:model2}
\end{align}
Each factor is controlled by a parameter, $\lambda$ for $\g$, $\omega$ for $\phi$ and $\sigma_H^2$ for $\nu_H$. We thus consider optimizing the long-term growth rate $\Lambda$ over $(\sigma_H^2,\lambda,\omega)$, for various values of $(a,\sigma_E^2,\sigma_D^2)$, using the expression for $\Lambda$ of the general model, with $\theta=1$, $\k=0$, $\rho=0$ (we consider $\l\geq 0$ and $\omega\geq 0$; see also Fig.~S2 for an alternative analysis where the optimization is performed over discrete values, $(\l,\omega)\in \{0,1\}^2$).\\

The result of a numerical optimization of $\Lambda$ are presented in Fig.~\ref{fig:QA}. They show that a feedback from phenotype to transmitted genotype is prevented ($\hat\omega=0$) in two limits: the limit of uncorrelated environments, $\tau_E\to 0$ ($a\to 0$), and the limit of deterministic environments $\sigma_E^2\to 0$. In the first limit, $\hat\l$ vanishes as well, hence the absence of feedback does not imply an isolated germ-line, but simply an absence of non-trivial heredity (for small $a$ but large $\sigma_E^2$, the solution $\hat\sigma_H^2>0$, $\hat\l=0$, $\hat\omega=0$ indicates that only noise is transmitted to the offsprings). The system of inheritance most reminiscent to Weismann's scenario is obtained in the limit of constant environments $(\tau_E,\sigma_E^2)\to(\infty,0)$ [$(a,\sigma_E^2)\to(1,0)$], suggesting that a separation of germ-plasm from soma is beneficial only for those aspects of the phenotype subject to non-fluctuating selective pressures ("housekeeping" genes, for instance).

\subsection{Where to acquire information?}\label{sec:whereinto}

In the two previous models, the role of the environment is confined to selection, and any new variation is introduced independently of the environmental state. Examples however abound of living organisms generating new traits that are correlated with the environment. It is thus well recognized that the current phenotype of an individual is not only affected by the genotype received from its parents, but also by the environment in which it develops; such a variation, which may be adaptive, is generally referred to as "plasticity". As already mentioned in the introduction, the question of whether environmentally induced variations can be transmitted to the progeny was, and still remains, a subject of hot debates. This possibility was central to Lamarck's theory of evolution, and is often referred to as "Lamarckism". Evolution by natural selection does not require it, but it does not exclude it. Several examples of environmentally induced traits have indeed been observed. Proving that such traits confer a selective advantage is generally delicate, but a particularly striking example is provided, for instance, by the bacterial immune system called CRISPR~\cite{Barrangou:2007bo}; this system relies on the insertion of phage-specific sequences into bacterial genomes and has been shown to protect against phage infection the bacteria that inherit them from their parent. This example implies a specific mechanism for incorporating and exploiting the environmental "signal" (here the presence of a particular strain of phage). The constraints to which the evolution of such mechanisms are subject are determining but potentially non-generic, and, in any case, difficult to model. Here, we consider a a thought experiment (or Gedankenexperiment), where we assume a mechanism for incorporating external signals, but question the way in which it is optimally "plugged in". This approach allows us, without discussing the mechanisms themselves, to compare the Darwinian and Lamarckian "modalities", and test the conjecture that each of them is tuned to a different type of  selective pressure~\cite{Koonin:2009dq}.\\

\begin{figure}[t]
\begin{center}
\includegraphics[width=\linewidth]{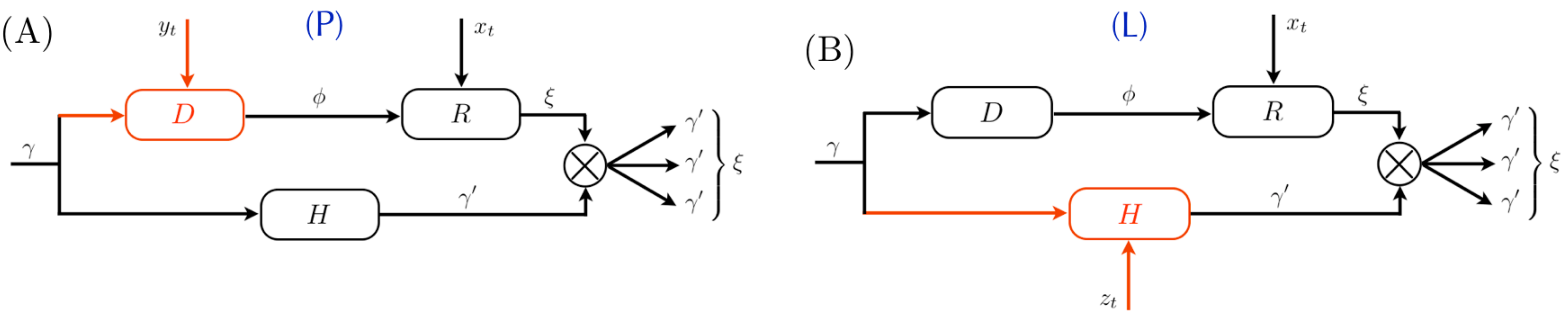}
\caption{Where to acquire information? This question is addressed by comparing two models in presence of the same information $y_t=x_t+b'_t$ with $b'_t\sim\N(0,\sq_I)$. {\bf (A)} Model P, described by Eqs~\eqref{eq:model3A}, where the information $y_t$ is incorporated to the phenotype and where the growth-rate $\Lambda$ is optimized with respect to the three parameters $\theta$, $\lambda$, $\rho$ which define $D(\phi|\g,y_t)$ and $H(\g'|\g)$ by the relations $\phi=\theta\g+\rho y_t+\nu_D$ and $\g'=\lambda\g+\nu_H$ with $\nu_D\sim\N(0,\sq_D)$ and $\nu_H\sim\N(0,\sq_H)$, and fixed $\sq_D$, $\sq_H$. {\bf (B)} Model L, described by Eqs~\eqref{eq:model3B}, where the information $y_t$ is incorporated to the transmitted genotype and where the growth-rate $\Lambda$ is optimized with respect to the three parameters $\theta$, $\lambda$, $\kappa$ which define $D(\phi|\g)$ and $H(\g'|\g,y_t)$ by the relations $\phi=\theta\g+\nu_D$ and $\g'=\lambda\g+\kappa y_t+\nu_H$ with $\nu_D\sim\N(0,\sq_D)$ and $\nu_H\sim\N(0,\sq_H)$, and fixed $\sq_D$, $\sq_H$.\label{fig:scheme3}}
\end{center}
\end{figure}

We thus compare two models, where the same information $y_t=x_t+b'_t$ with $b'_t\sim\N(0,\sq_I)$ is available, but where it is either processed at the phenotypic level (model P for "plasticity", for which $\sq_I=\sq_{Ip}$, Fig.~\ref{fig:scheme3}A), or at the genotypic level (model L for "Lamarckism", for which $\sq_I=\sq_{I\ell}$, Fig.~\ref{fig:scheme3}B). Formally, model P is described by
\begin{align}
\g'&=\lambda\g+\nu_H,\qquad\qquad\quad\ \nu_H\sim\N(0,\sq_H),\nonumber\\
\phi&=\theta\g+\rho y_t+\nu_D,\qquad\quad\nu_D\sim\N(0,\sq_D),\label{eq:model3A}
\end{align}
thus corresponding to the general model with $\omega=0$ and $\kappa=0$, while model L is described by
\begin{align}
\g'&=\lambda\g+\k y_t+\nu_H,\qquad\nu_H\sim\N(0,\sq_H),\nonumber\\
\phi&=\theta\g+\nu_D,\qquad\qquad\quad\nu_D\sim\N(0,\sq_D),\label{eq:model3B}
\end{align}
thus corresponding to the general model with $\omega=0$ and $\rho=0$. We consider optimizing the long-term growth rate $\Lambda$ over the parameters that control the contributions of each factor: $(\theta,\lambda,\rho)$ for model P, and $(\theta,\lambda,\kappa)$ for model L (considering here again only positive values of these parameters).\\

\begin{figure}[t]
\begin{center}
\includegraphics[width=.8\linewidth]{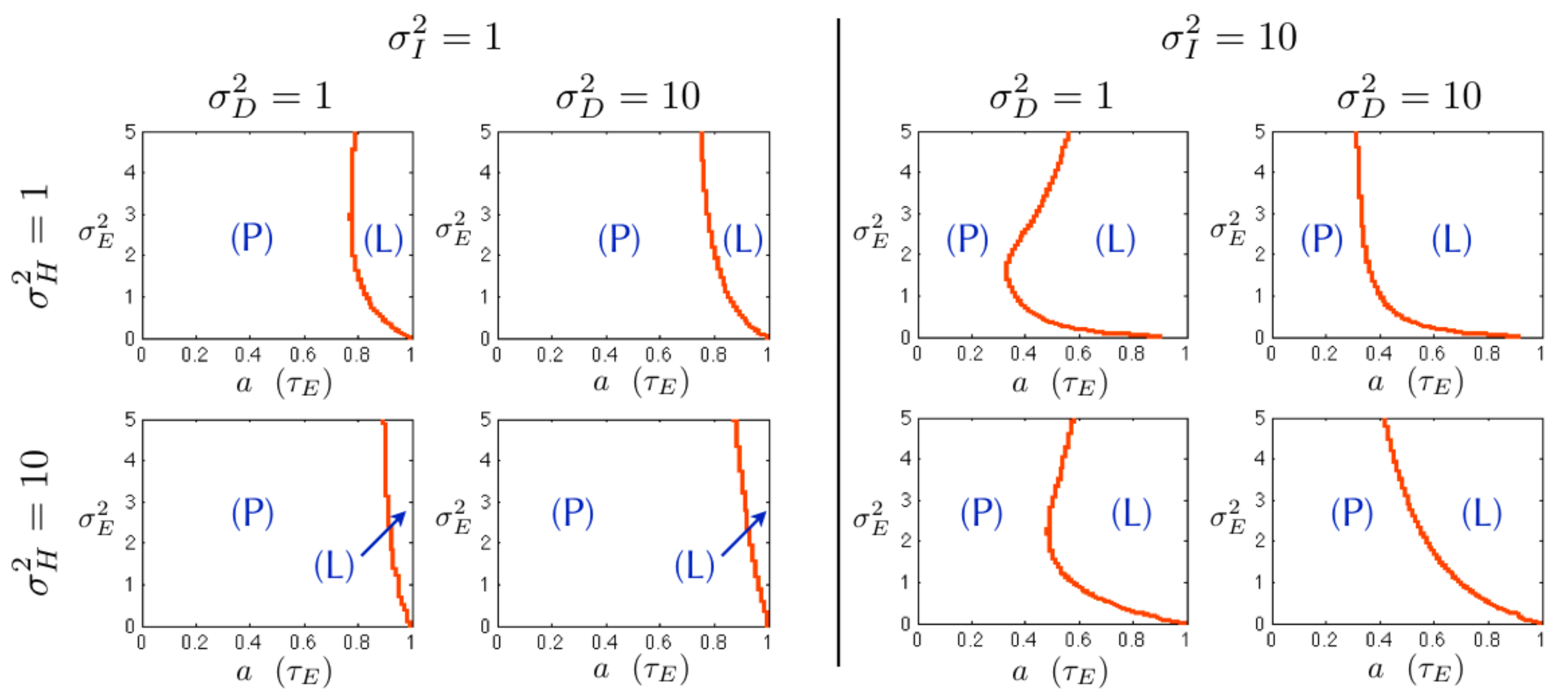}
\caption{Where to acquire information? -- Boundary between $\hat\Lambda(\kappa=0)>\hat\Lambda(\rho=0)$, when acquiring information at the phenotypic level is more beneficial, indicated by (P), and $\hat\Lambda(\kappa=0)<\hat\Lambda(\rho=0)$, when acquiring information at the genotypic level is more beneficial, indicated by (L), for all 8 combinations of $(\sigma_H^2,\sigma_D^2,\sigma_I^2)\in\{1,10\}^3$. $\sigma_I^2$ represents $\sq_{Ip}$ for model P and $\sq_{I\ell}$ for model L.}\label{fig:Q3}
\end{center}
\end{figure}

The results of a comparison between the two models are shown in Fig.~\ref{fig:Q3}, where, as a function of $(a,\sigma_E^2)$ and for different values of the fixed parameters $\sigma_I^2,\sigma_H^2,\sigma_D^2$, we present which of the two models, P or L, yields the highest growth rate. The main controlling parameter appears to be the correlation $a$ (or equivalently $\tau_E$) of the environmental fluctuations, with the Lamarckian modality systematically becoming more favorable when this correlation is large, in line with the intuition that transmitting acquired information is beneficial when the selective pressure experienced by the offspring is sufficiently similar to that experienced by the parents. Note that this simple conclusion conceals in fact a much richer diversity of strategies, revealed by considering the values of the parameters optimizing the two models (Fig.~S\ref{fig:Q3S}). 

\section{Connections}

Of the many studies that share part of our intents or methods, two lines of work stand out: (i) a different approach, based on Price's equation, has been proposed with the same goal of uniting the various forms of inheritance within a common mathematical framework~\cite{Helantera:2010tf,Day:2011fn}; (ii) a different problem, pertaining to control in engineering, has been solved using a closely related mathematical framework~\cite{Kalman:1960tn}. We discuss here the relations between our model and these two lines of work.

\subsection{Link to Price equation}\label{sec:price}

Price equation is a general formula, applicable to any model of population dynamics, which uses a covariance formalism to express the change in the mean value of a trait between successive generations~\cite{Price:1970vw}. Being a mathematical identity, it necessarily holds true. Its virtue is to provide a decomposition of evolutionary change that can illuminate its origins; initially derived for models of cooperation, it has contributed to clarify the nature of group/kin selection~\cite{Price:1972wv}. Applied in many other contexts, it has also been used as a general mathematical framework for studying the different possible modes of inheritance~\cite{Helantera:2010tf,Day:2011fn}. As for any other model of population dynamics, a Price equation can be written for our model. For the simple model of Sec.~\ref{sec:simple}, it thus takes the form 
\beq\label{eq:price}
\langle\xi\rangle_t\ \Delta_t\langle\g\rangle=\textrm{cov}_t(\xi,\g)+\langle \xi\Delta\g\rangle_t,
\eeq
where, following the usual notation for Price equation, $\langle \xi\rangle_t=W_t$ is the "instantaneous fitness", $\langle \g\rangle_t=m_t$ is the mean value of the attribute $\g$, $\Delta_t\langle\g\rangle =m_{t+1}-m_t$ is the change of its mean between two generations, $\textrm{cov}_t(\xi,\g)=\langle\xi\g\rangle_t-\langle\xi\rangle_t\langle\g\rangle_t$ is the covariance between the attribute and the instantaneous fitness, and 
\beq
\Delta \g=\g\int\ud \g'\ \left[R(\xi|\g',x_t)\ H(\g|\g')\ n_t(\g')-R(\xi|\g,x_t)\ H(\g'|\g)\ n_t(\g)\right].
\eeq
(see derivation in Suppl. Info.). Eq.~\eqref{eq:price} can be interpreted as a decomposition of the evolutionary change into a contribution due to selection, $\textrm{cov}_t(\xi,\g)$, and a contribution from mutations, $\langle \xi\Delta\g\rangle_t$; the latter quantity indeed vanishes when transmission is error-free, i.e., when $H(\g'|\g)=\delta(\g'-\g)$.\\

Price equation is limited to a short-term description of the dynamics: as it considers only the mean values of traits, it cannot be iterated to describe changes in subsequent generations; this indeed requires the full distribution $n_t(\g)$. As illustrated in the previous sections, systems of inheritance may, however, have qualitatively different implications in environments with different statistical structures. Only a long-term analysis of the dynamics of a population can thus fully reveal their evolutionary properties. In our approach, this is achieved by coupling Price equation, the recursion over the mean $m_t$ of the trait, Eq.~\eqref{eq:rec1}, with a recursion over the variance $\sq_t$ of the trait, Eq.~\eqref{eq:rec2}, and by considering their asymptotic properties, which consists of a fixed point for $\sq_t$ and a stationary distribution for $m_t$. Within the Gaussian assumptions that define our model, these two quantities are sufficient to fully capture the population dynamics. More importantly, our formalism involves a central quantity not present in covariance formalisms, the Lyapunov exponent $\Lambda$. Because this quantity decides the eventual fate of two competing populations (see Sec.~\ref{sec:whyL}), it allows us to associate with each scheme of inheritance an adaptive value, and thus to derive the conditions under which a given scheme confers a selective advantage over the others.

\subsection{Link to the Kalman filter}

At the core of adaptation is the problem of anticipating the next state of the environment. From this perspective, different systems of inheritance can be considered as different ways to share with the current generation the "knowledge" accumulated, through natural selection, by the previous generations. In a stochastically fluctuating environment, no system of inheritance can, however, perform better than direct sensing of the present environment by the individual that experiences it. Sensors, when not altogether absent, are generically imperfect. The individual with such an imperfect access to its current environment thus faces a dilemma: it has to arbitrate between two valuable but unreliable sources of information, the "germ-plasm" $\g$ inherited from the parent, and the cues $y_t$ or $z_t$ that it gained from its own perception. Interestingly, the very same dilemma is encountered in various problems of engineering, such as for instance the automatic guidance of aircrafts, where decisions must also be made based on two potentially conflicting sources: the past states of the system, and the signals from the sensors. A classical algorithm for solving this problem is the Kalman filter~\cite{Kalman:1960tn}. Maybe not surprisingly, it involves the same essential mathematical ingredients that make our model solvable: linearity and Gaussianity. We present here a limit case of our model where the two approaches formally coincide, thus revealing that the scope of the concepts of inheritance extends beyond the study of biological organisms.\\

Control in engineering typically involves a single system, rather than a population of diverse individuals. We thus obtain a formal analogy with the Kalman filter only when the population is perfectly homogeneous, and described by a single "state" $\g$. This corresponds in our model to a limit where no developmental noise is present, $\phi=\g$ (i.e., $\theta=1$, $\rho=0$, $\sq_D=0$), and where the environment is perfectly selective, $\sq_S= 0$, so that each and every individual has a common $\g=x_t$. In this limit case, the heredity kernel $\hat H$ that optimizes $\Lambda$ can be computed exactly for any stationary Markovian environmental process (not necessarily Gaussian).\\

First, assuming that no external information is available, with a model described by $A(\xi,\g'|\g,x_t)=R(\xi|\g',x_t)H(\g'|\g)$ where $\sum_\xi\xi\ R(\xi|\g,x_t)=e^{r_{\rm max}}\delta(\g-x_t)$, and denoting by $P(x_{t+1}|x_t)$ the stochastic kernel defining the environmental process, we have
\beq
\Lambda=r_{\rm max}+\lim_{t\to\infty}\frac{1}{t}\E[P(x_{t+1}|x_t)\ln H(x_{t+1}|x_t)]=r_{\rm max}-\mathcal{H}(P)-\mathcal{D}(P\|H),
\eeq
where $\mathcal{H}(P)$ denotes the entropy rate of the process $P$, and $\mathcal{D}(P\|H)$ the rate of relative entropy\footnote{They are defined by $\mathcal{H}(P)=-\lim_{t\to\infty}\E[P(x_{t+1}|x_t)\ln P(x_{t+1}|x_t)]/t$ and $\mathcal{D}(P\|H)=\lim_{t\to\infty}\E[P(x_{t+1}|x_t)\ln P(x_{t+1}|x_t)/H(x_{t+1}|x_t)]/t$ ~\cite{CoverThomas91}.} of $P$ with respect to $H$~\cite{Rivoire:2011ue}. The later verifies $\mathcal{D}(P\|H)\geq 0$, with $\mathcal{D}(P\|H)= 0$ if and only if $P=H$ \cite{CoverThomas91}. The growth rate $\Lambda$ is therefore minimal for $\hat H=P$, an instance of the so-called proportional betting strategy~\cite{KellyJr:1956un}, which consists in matching the stochasticity of the environment. In particular, when taking a Gaussian environment with $P(x_{t+1}|x_t)= G_{\sq_X}(x_{t+1}-ax_t)$, we obtain $\hat H(\g'|\g)=G_{\sq_X}(\g'-a\g)$, corresponding to $\hat\l=a$ and $\hat\s^2_H=\sq_X$.\\

Assuming now that some information $y_t$ is available, which is derived from $x_t$ as\footnote{We previously assumed $c_I=1$; $\k$ and $\sq_I$ can indeed always be rescaled to be in this case. We introduce here $c_I$ only to make the correspondence with the Kalman filter where this rescaling is generally not assumed.} $y_t=c_Ix_t+b'_t$ with $b'_t\sim\N(0,\sq_I)$, we can 
extend this result to a model with $A(\xi,\g'|\g,x_t)=R(\xi|\g',x_t)H(\g'|\g,y_t)$. This model, which corresponds to a continuous version of the model studied in Ref.~\cite{Rivoire:2011ue}, was previously analyzed in Ref.~\cite{Haccou:1995tf}. In the limit where $\sum_\xi\xi\ R(\xi|\g,x_t)=e^{r_{\rm max}}\delta(\g-x_t)$, the growth rate $\Lambda$ is optimized by following a Bayesian strategy
\beq
\hat H(\g'|\g,y_t)=\P_{X_{t+1}|X_t,Y_t}(\g'|\g,y_t),
\eeq
where $\P_{X_{t+1}|X_t,Y_t}(x_{t+1}|x_t,y_t)$ denotes the conditional probability of having the environmental random variable $X_{t+1}$ taking the value $x_{t+1}$ at time $t+1$ given that it took the value $x_t$ at time $t$, and that the population observed $y_t$ at this time. 
With a Gaussian environment and Gaussian noisy channel, $\P_{X_{t+1}|X_t}(x_{t+1}|x_t)=G_{\sq_X}(x_{t+1}-ax_t)$ and $\P_{Y_t|X_t}(y_t|x_t)=G_{\sq_I}(y_t-c_Ix_t)$, this yields\footnote{In this case, it is thus proved that the optimal form of heritability kernel $H$ is Gaussian.}
\beq
\hat H(\g'|\g,y_t)=G_{((\s_Ic_I)^{-2}+\s_X^{-2})^{-1}}\left(\g'-\frac{a\s_I^2}{\s_I^2+c_I^2\s_X^2}\g-\frac{c_I\s_X^2}{\s_I^2+c_I^2\s_X^2}y_t\right).
\eeq
The two coefficients $\hat\l=a\s_I^2/(\s_I^2+c_I^2\s_X^2)$ and $\hat\k=c_I\s_X^2/(\s_I^2+c_I^2\s_X^2)$ correspond exactly to the gains for the Kalman filter~\cite{Kalman:1960tn}; in this context, they prescribe how the previous state $\g$ of the system and the newly acquired information $y_t$ must be linearly combined to optimally define the subsequent state $\g'$ of the system.\\

The formal correspondence with the solution to the Kalman filter holds only in the limit of infinite selectivity $\sq_S\to 0$, where the population is perfectly homogeneous at every time. This limit, where the optimal scheme for processing information follows the Bayesian principles, is also the limit in which the value of the information $y_t$ can be quantified by the usual concepts of information theory~\cite{Rivoire:2011ue}. We may thus view our model as a generalization of the problem of stochastic control encountered in engineering by incorporating biological features that are absent in this context, notably a diverse and growing population, and a distinction between genotype and phenotype. Reciprocally, the Kalman filter has been extended along several lines since its original formulation~\cite{Simon06}, and the mathematical formalisms thus developed may suggest ways along which generalizations of our model could be analyzed.

\section{Conclusion}

Classical models of population genetics take the mechanisms for generating and transmitting new traits as given. Several previous studies have extended these models to analyze how the mechanisms of inheritance may themselves evolve, starting from works on the evolution of mutation rates~\cite{Kumura67} and including, among several other examples, studies of maternal effects~\cite{Kirkpatrick:1989uw}, non-genetic inheritance~\cite{Lachmann:1996uq}, plasticity and memory~\cite{Pal:1998th}, and relationship quantitative trait loci~\cite{Pavlicev:2011iv}. Here, we proposed a simple model to compare the adaptive value of different schemes for generating and transmitting variations in populations. Its analysis indicates that different modalities of inheritance are favored depending on the statistical structure of the fluctuations of the environment. For an organism with various traits, each potentially subject to a different selective pressure, this analysis suggests that multiple inheritance systems operating in parallel may be selected for, consistently with observations.\\ 

While general, our model certainly does not encompass the full diversity of possible modes of inheritance. It may still be extended along several lines while retaining its analytical tractability. For instance, rather than combining the inherited and the acquired information into a single attribute, it could include two channels of transmission, one for the germ-line and another for somatic elements, as for instance in Ref.~\cite{Kirkpatrick:1989uw}; since Gaussian formulae extend to multidimensional variables, the model remains indeed solvable when considering the generation and transmission of multiple traits. It can similarly be extended to account for multiple time scales, for instance by introducing a temporal delay between the developmental stage and the time of reproduction (formally $y_t= x_{t-\tau}$, for $0<\tau<1$). Different environmental processes can also be analyzed, which do not need to be Gaussian for the model to be solvable; e.g., adding a linearly varying component to the environment ($x'_t=x_t+k t$) provides a framework for studying the implications of different modes of adaptations to the survival of a population facing a directed change of selective pressure~\cite{Chevin:2010bg}.\\

The model is abstracted from material implementations, and, in particular, does not refer to the genetic or non-genetic nature of what is transmitted. This abstraction confers to our model its generality, but also underlies one of its main limitations: it cannot account for the constraints and costs that the evolution of any specific mechanism for generating and transmitting variations must face. The model can certainly be extended to include such costs, both constitutive (attached to the mechanisms) or inductive (stemming from their usage), but only at the price of introducing new, {\it ad hoc} parameters. This limitation for example prevents us from making a meaningful comparison of the relative benefit of "selection" versus "instruction", since selection results automatically from the interaction with the environment, while a specific mechanism is needed to acquire and incorporate environmental signals. More generally, the model does not account for the fact that a reliable hereditary mechanism must precede the evolution of a Lamarckian mechanism, if this  mechanism is to be faithfully transmitted.\\

Despite these limitations, we hope that our approach may be of value for providing theoretical limits to the evolution of systems of inheritance, in the spirit of the theoretical limits that Shannon derived for the communication of signals over noisy channels, after similarly abstracting from practical costs and constraints~\cite{Shannon:1948wk}. As shown in our previous work~\cite{Rivoire:2011ue}, the similarity between the two problems extends beyond the mere analogy: the fundamental quantities of information theory are recovered as a limit of our model. We exposed here, in the same limit, another formal analogy, with the solution to the Kalman filter used in stochastic control~\cite{Kalman:1960tn}. The model presented in this paper thus provides a general, analytically tractable framework for clarifying and unifying common issues and concepts in population genetics, information theory and stochastic control, which may contribute to stimulate further crossbreeding between these disciplines.

\acknowledgments

We thank B. Chazelle, B. Houchmandzadeh, D. Huse, D. Jordan and S. Kuehn for comments. This work was supported by ANR grant CoevolInterProt (to O.R.).

\bibliographystyle{apsrev}


\newpage

\appendix
\setcounter{figure}{0}

\noindent {\Large SUPPLEMENTARY INFORMATION}

\section{Analytical formula for the long-term growth rate $\L$}

We detail here the derivation of the analytical formula for the growth rate $\L$ of the general model of Sec.~\ref{sec:general}, Eq.~\eqref{eq:general}. The derivation relies on the stability of Gaussian functions under multiplication and convolution. Denoting by $G_{\sq}(x)$ a generic Gausian function, i.e.,
\beq
G_{\sq}(x)\equiv\frac{1}{(2\pi\sq)^{1/2}}\exp\left({-\frac{x^2}{2\sq}}\right),
\eeq
the property of stability under multiplication corresponds to the identity 
\beq\label{eq:id1}
\begin{split}
&G_{\s_1^2}(\l_1x-m_1)G_{\s_2^2}(\l_2 x-m_2)\\
&=G_{(\l_1^2\s_1^{-2}+\l_2^2\s_2^{-2})^{-1}}\left(x-\frac{\l_1\s_2^2}{\l_2^2\s_1^2+\l_1^2\s_2^2}m_1-\frac{\l_2\s_1^2}{\l_2^2\s_1^2+\l_1^2\s_2^2}m_2\right)G_{\l_2^2\s_1^2+\l_1^2\s_2^2}(\l_2m_1-\l_1m_2).
\end{split}
\eeq
This identity implies the stability under convolution:
\beq\label{eq:id2}
\int \ud x\ G_{\s_1^2}(\l_1 x-m_1)G_{\s_2^2}(\l_2 x-m_2)=G_{(\l_2\s_1)^2+(\l_1\s_2)^2}(\l_2 m_1-\l_1m_2).
\eeq

\subsection{Particular model}\label{sec:rec}

We first treat a simpler model, to which the general model will correspond once properly parametrized. This particular model, obtained from the general model by setting  $\sq_D=0$, $\theta=1$, $\rho=0$, $\omega=0$, is itself a slight generalization of the model of Sec.~\ref{sec:simple}. It is defined by $A(\xi,\g'|\g,x_t)=R(\xi|\g,x_t)H(\g'|\g,x_t)$, with
\beq
\sum_\xi \xi\ R(\xi|\g,x_t)=\exp\left(r_{\rm max}-\frac{(\g-x_t)^2}{2\sq_S}\right),\qquad H(\g'|\g,x_t)=G_{\sq_H}(\g'-\l\g-\k x_t).
\eeq
Considering an environment following $x_{t+1}=ax_t+b_t$ with $b_t\sim\N(0,\sq_X)$, and ignoring the trivial factor $r_{\rm max}$ ($\L$ with $r_{\rm max}>0$ is obtained as $\L=\L(r_{\rm max}=0)+r_{\rm max}$), this model has in total 6 parameters: $a$, $\sq_X$, $\sq_S$, $\sq_H$, $\l$, $\k$.\\

We start from the recursion for the composition  $n_t(\g)$ of the population, Eq.~\eqref{eq:meanfield}, which we write under the assumption that $n_t(\g)$ is Gaussian, $n_t(\g)=G_{\sq_t}(\g-m_t)$, 
\beq
W_t\ n_{t+1}(\g')=(2\pi\sq_S)^{1/2}\int\ud\g\ G_{\sq_S}(\g-x_t)G_{\sq_H}(\g'-\l\g-\k x_t)G_{\sq_t}(\g-m_t).
\eeq
We first use Eq.~\eqref{eq:id1} to obtain
\begin{eqnarray}
W_t&=&(2\pi\sq_S)^{1/2}G_{\sq_S+\sq_t}(m_t-x_t),\label{eq:W}\\
n_{t+1}(\g')&=&\int\ud\g\ G_{\sq_H}(\g'-\l\g-\k x_t)G_{(\s_R^{-2}+\s_t^{-2})^{-1}}\left(\g-\frac{\sq_t}{\sq_S+\sq_t}x_t-\frac{\sq_S}{\sq_S+\sq_t}m_t\right),
\end{eqnarray}
and then Eq.~\eqref{eq:id2} to identify the mean $m_{t+1}$ and variance $\sq_{t+1}$ of $n_{t+1}(\g')=G_{\sq_{t+1}}(\g'-m_{t+1})$,
\bea
m_{t+1}&=\l\frac{\sq_S}{\sq_S+\sq_t}m_t+\left(\l\frac{\sq_t}{\sq_S+\sq_t}+\k\right) x_t,\label{eq:recmt}\\
\sq_{t+1}&=\l^2(\s_S^{-2}+\s_t^{-2})^{-1}+\sq_H.\label{eq:recst}
\eea
Eq.~\eqref{eq:W} can also be written
\beq\label{eq:lnWtparticular}
\ln W_t=\frac{1}{2}\ln\left(\frac{\sq_S}{\sq_S+\sq_t}\right)-\frac{1}{2}\frac{(m_t-x_t)^2}{\sq_S+\sq_t},
\eeq
from which the long-term growth rate $\Lambda$ is derived as $\Lambda=\lim_{t\to\infty}\E [\ln W_t]$. Taking $\l=1$ and $\k=0$, and introducing the notation $h_t^2\equiv\sq_t/(\sq_t+\sq_S)$, this leads to the expression given in the main text for the simple model, Eq.~\eqref{eq:lnW}.\\

In the limit $t\to\infty$, the recursion for $\sq_t$, Eq.~\eqref{eq:recst}, leads a fixed point $\sq_\infty$ that is solution to
\beq\label{eq:sqinfty}
\sq_\infty=\l^2(\s_\infty^{-2}+\s_S^{-2})^{-1}+\sq_H.
\eeq
For large $t$, we can therefore approximate the recursion for $m_t$, Eq.~\eqref{eq:recmt}, by 
\beq\label{eq:mtgen}
m_{t+1}-x_{t+1}=\a(m_t-x_t)-(x_{t+1}-\left(\l+\k)x_t\right),
\eeq
where $\alpha$, defined by
\beq
\a\equiv\frac{\l\sq_S}{\sq_S+\sq_\infty},
\eeq
is, after solving Eq.~\eqref{eq:sqinfty}, found to be
\beq\label{eq:alphal}
\alpha=\frac{2\lambda}{1+\lambda^2+\beta+\left((1-\lambda^2-\beta)^2+4\beta\right)^{1/2}},\qquad\textrm{with}\qquad \beta\equiv\frac{\sq_H}{\sq_S}.
\eeq
With these notations, the long-term growth rate $\Lambda=\lim_{t\to\infty}\E [\ln W_t]$ can be written
\beq\label{eq:evload}
\Lambda=\frac{1}{2}\ln\left(\frac{\alpha}{\lambda}\right)-\frac{1}{2}\frac{\alpha}{\lambda}\lim_{t\to\infty}\frac{\E[(m_t-x_t)^2]}{\sq_S}.
\eeq

Completing the solution requires estimating $\lim_{t\to\infty}\E[(m_t-x_t)^2]$. To this end, we first rewrite Eq.~\eqref{eq:mtgen} as
\beq
m_{t+1}-x_{t+1}=\a(m_t-x_t)+\epsilon x_t-b_t,\quad\textrm{with}\quad \epsilon\equiv\l+\k-a,
\eeq
which implies
\beq
m_{t+1}-x_{t+1}=\sum_{k=0}^t \a^{t-k}(\epsilon x_k-b_k).
\eeq
Using the relation $x_k=\sum_{n=0}^{k-1}a^{k-1-n}b_n$, which follows from the recursion $x_k=ax_{k-1}+b_{k-1}$, we have
\beq
\sum_{k=0}^t \a^{t-k}x_k=\sum_{k=0}^{t-1}\sum_{n=0}^{t-k-1}\a^{t-n-k-1}a^nb_k=\sum_{k=0}^{t-1}\frac{1-(a\a^{-1})^{t-k}}{1-a\a^{-1}}\a^{t-(k+1)}b_k=\sum_{k=0}^{t-1}\frac{\a^{t-k}-a^{t-k}}{\a-a}b_k,
\eeq
where the last sum can be extended up to $k=t$. We thus get
\beq\label{eq:mxbk}
m_{t+1}-x_{t+1}=\frac{1}{\alpha-a}\sum_{k=0}^t [(\epsilon-\alpha+a)\a^{t-k}-\epsilon a^{t-k}]b_k.
\eeq
Since the $b_k$'s are independent and identically distributed random variables with variance $\sq_X$, we have
\beq
\E[(m_{t+1}-x_{t+1})^2]=\frac{1}{(\alpha-a)^2}\sum_{k=0}^t [(\epsilon-\alpha+a)\a^{t-k}-\epsilon a^{t-k}]^2\sq_X.
\eeq
Summing the geometric series and taking the limit $t\to\infty$ then leads to
\beq
\lim_{t\to\infty}\E[(m_t-x_t)^2]=\frac{1}{(\a-a)^2}\left(\frac{(\epsilon-\alpha+a)^2}{1-\a^2}-\frac{2\epsilon(\epsilon-\alpha+a)}{1-\a a}+\frac{\epsilon^2}{1-a^2}\right)\sq_X,
\eeq
or, equivalently,
\beq
\lim_{t\to\infty}\E[(m_t-x_t)^2]=\frac{(1+a\a)(1+(\epsilon+a)^2)-2(a+\alpha)(\epsilon+a)}{(1-a^2)(1-\a^2)(1-a\a)}\sq_X.
\eeq
Finally, replacing $\lim_{t\to\infty}\E[(m_t-x_t)^2]$ with this expression and invoking the relations $1/(\sq_\infty+\sq_S)=\a/(\l\sq_S)$ and $\epsilon=\l+\k-a$ yields
\beq\label{eq:Lparticular}
\Lambda=\frac{1}{2}\ln\left(\frac{\alpha}{\l}\right)-\frac{\alpha}{2\l}\frac{(1+a\alpha)(1+(\l+\k)^2)-2(a+\alpha)(\l+\k)}{(1-a^2)(1-\alpha^2)(1-a\alpha)}\frac{\sq_X}{\sq_S},
\eeq
which, together with Eq.~\eqref{eq:alphal}, provides an analytical formula for the long-term growth rate  $\Lambda$ of this particular model. With $\l=1$ and $\k=0$, this formula corresponds to the expression for $\Lambda$ given in the main text for the simple model, Eq.~\eqref{eq:Lsimple}.

\subsection{General model}\label{sec:sol}

The general model is defined in Sec.~\ref{sec:general} by $A(\xi,\g'|\g,x_t)=\int\ud\phi\ R(\xi|\g,\phi,x_t)D(\phi|\g,y_t)H(\g'|\g,\phi,z_t)$, with 
\beq
\sum_\xi \xi\ R(\xi|\g,x_t)=(2\pi\sq_S)^{1/2} G_{\sq_S}(\g-x_t),\quad D(\phi|\g,y_t)=G_{\sq_D}(\g'-\theta\g-\rho y_t),\quad H(\g'|\g,\phi,z_t)=G_{\sq_H}(\g'-\l\g-\omega\phi-\k z_t).
\eeq
(we assume here $r_{\rm max}=0$). All the environmental variables, represented by Roman letters, are defined independently of the dynamics of the population, by $x_{t+1}=ax_t+b_t$, $y_t=x_t+b'_t$, and $z_t=x_t+b''_t$ with $b_t\sim\N(0,\sq_X)$, $b'_t\sim\N(0,\sq_{Ip})$ and $b''_t\sim\N(0,\sq_{I\ell})$.\\

Under the assumption that the attribute $\g$ is normally distributed in the population, $n_t(\g)=G_{\sq_t}(\g-m_t)$, the basic recursion defining the model is
\beq
W_t\ n_{t+1}(\g')=(2\pi\sq_S)^{1/2}\int\ud\g\ud\phi\  G_{\sq_S}(\phi-x_t)G_{\sq_D}(\phi-\theta\g-\rho y_t)G_{\sq_H}(\g'-\l\g-\omega\phi-\k z_t)G_{\sq_t}(\g-m_t).
\eeq

The first steps that we take for solving this general model mirror those taken to solve the particular model. 
Using Eq.~\eqref{eq:id1}, $G_{\sq_S}(\phi-x_t)G_{\sq_D}(\phi-\theta\g-\rho y_t)$ is rewritten as $P(\phi,\g)X(\g)$ with
\beq
P(\phi,\g)=G_{(\s_D^{-2}+\s_S^{-2})^{-1}}\left(\phi-\frac{\sq_D}{\sq_D+\sq_S}x_t-\frac{\sq_S}{\sq_D+\sq_S}(\theta\g+\rho y_t)\right),\qquad X(\g)=G_{\sq_D+\sq_S}\left(\theta\g-(x_t-\rho y_t)\right).
\eeq
The integration over $\phi$ in $W_t n_{t+1}(\g')=(2\pi\sq_S)^{1/2}\int\ud\g\ud\phi\ P(\phi,\g)X(\g)H(\g'|\g,\phi,z_t)G_{\sq_t}(\g-m_t)$ involves only $Y(\g,\g')=\int \ud\phi\ P(\phi,\g)H(\g'|\g,\phi,z_t)$ with
\beq
Y(\g,\g')=G_{\omega^2(\s_D^{-2}+\s_S^{-2})^{-1}+\sq_H}\left(\left(\l+\frac{\sq_S}{\sq_D+\sq_S}\omega\theta\right)\g-\g'+\frac{\sq_D}{\sq_D+\sq_S}\omega x_t+\frac{\sq_S}{\sq_D+\sq_S}\omega\rho y_t+\k z_t\right),
\eeq
so that $W_t n_{t+1}(\g')=(2\pi\sq_S)^{1/2}\int\ud\g\  X(\g)Y(\g,\g')G_{\sq_t}(\g-m_t)$. By rewriting $(2\pi\sq_S)^{1/2}X(\g)G_{\sq_t}(\g-m_t)$ as $W_tZ(\g)$ we obtain
\beq
Z(\g)=G_{(\theta^2(\sq_D+\sq_S)^{-1}+\s_t^{-2})^{-1}}\left(\g-\frac{\sq_t}{\sq_D+\sq_S+\theta^2\sq_t}\theta(x_t-\rho y_t)-\frac{\sq_D+\sq_S}{\sq_D+\sq_S+\theta^2\sq_t}m_t\right),
\eeq
and
\beq
W_t=(2\pi\sq_S)^{1/2}G_{\sq_D+\sq_S+\theta^2\sq_t}(\theta m_t-(x_t-\rho y_t)),
\eeq
i.e.,
\beq
\ln W_t=\frac{1}{2}\ln\left(\frac{\sq_S}{\sq_D+\sq_S+\theta^2\sq_t}\right)-\frac{(\theta m_t-(x_t-\rho y_t))^2}{2(\sq_D+\sq_S+\theta^2\sq_t)}.
\eeq
Finally, $ n_{t+1}(\g')=\int\ud\g\  Y(\g,\g')Z(\g)=G_{\sq_{t+1}}(\g'-m_{t+1})$, with
\beq
\sq_{t+1}=\omega^2(\s_D^{-2}+\s_S^{-2})^{-1}+\sq_H+\left(\l+\frac{\sq_S}{\sq_D+\sq_S}\omega\theta\right)^2(\theta^2(\sq_D+\sq_S)^{-1}+\s_t^{-2})^{-1},
\eeq
and
\beq
\begin{split}
m_{t+1}=&\left(\l+\frac{\sq_S}{\sq_D+\sq_S}\omega\theta\right)\frac{\sq_D+\sq_S}{\sq_D+\sq_S+\theta^2\sq_t}m_t+\left(\frac{\sq_D}{\sq_D+\sq_S}\omega+\left(\l+\frac{\sq_S}{\sq_D+\sq_S}\omega\theta\right)\frac{\sq_t}{\sq_D+\sq_S+\theta^2\sq_t}\theta\right) x_t\\
&+\left(\frac{\sq_S}{\sq_D+\sq_S}\omega-\left(\l+\frac{\sq_S}{\sq_D+\sq_S}\omega\theta\right)\frac{\sq_t}{\sq_D+\sq_S+\theta^2\sq_t}\theta\right) \rho y_t+\k z_t.
\end{split}
\eeq

At this stage, it is convenient to introduce new variables that effectively map these equations to the same equations found for the particular model of the previous section. By defining
\beq
\tilde\s_S^2\equiv\frac{\sq_D+\sq_S}{\theta^2},\qquad\tilde\s_H^2\equiv\sq_H+\omega^2(\s_D^{-2}+\s_S^{-2})^{-1},\qquad\tilde\l\equiv\l+\omega\theta\frac{\sq_S}{\sq_D+\sq_S},
\eeq
we thus have
\beq
\sq_{t+1}=\tilde\l^2(\s_t^{-2}+\tilde\s_S^{-2})^{-1}+\tilde\s_H^2,
\eeq
which is formally identical to Eq.~\eqref{eq:recst}. Similarly, with
\beq
\tilde x_t\equiv\theta^{-1}(x_t-\rho y_t)
\eeq
we obtain
\beq
\ln W_t=\frac{1}{2}\ln\left(\frac{\tilde\s_S^2}{\tilde\s_S^2+\sq_t}\right)-\frac{(m_t-\tilde x_t)^2}{2(\tilde\s_S^2+\sq_t)}+\frac{1}{2}\ln\left(\frac{\s_S^2}{\s_S^2+\s_D^2}\right),
\eeq
which, up to an additive constant (which could be absorbed by redefining $r_{\rm max}$), is formally similar to Eq.~\eqref{eq:lnWtparticular}. Finally, we can also write the recursion for $m_t$ as
\beq
m_{t+1}=\tilde\l\frac{\tilde\s_S^2}{\tilde\s_S^2+\sq_t} m_t+\left(\tilde\l\frac{\sq_t}{\tilde\s_S^2+\sq_t}+\tilde \k\right)\tilde x_t+c_t.
\eeq
with
\beq
\tilde\k\equiv\left(\frac{\sq_D+\rho \sq_S}{\sq_D+\sq_S}\omega+\k \right)\frac{\theta}{1-\rho}.
\eeq
and
\beq
c_t\equiv\left(\frac{\rho(\omega+\k)}{1-\rho }\right)b'_t+\k b''_t.
\eeq
Here again, we obtain, up to the addition of the random variable $c_t$, a relation for $m_t$ that is formally similar to the one satisfied by the particular model, Eq.~\eqref{eq:recst}.\\

Given these correspondences, we can derive the long-term growth rate $\L$ of the general model from the expression found for the particular model, Eq.~\eqref{eq:Lparticular}, without repeating the intermediate steps. This leads to the expression of $\L$ given in the main text by Eq.~\eqref{eq:general}. 

\section{Optimization of $\L$}

\subsection{Simple model of Sec.~\ref{sec:simple}}

\begin{figure}[h]
\renewcommand{\figurename}{FIG. S}
\begin{center}
\includegraphics[width=.6\linewidth]{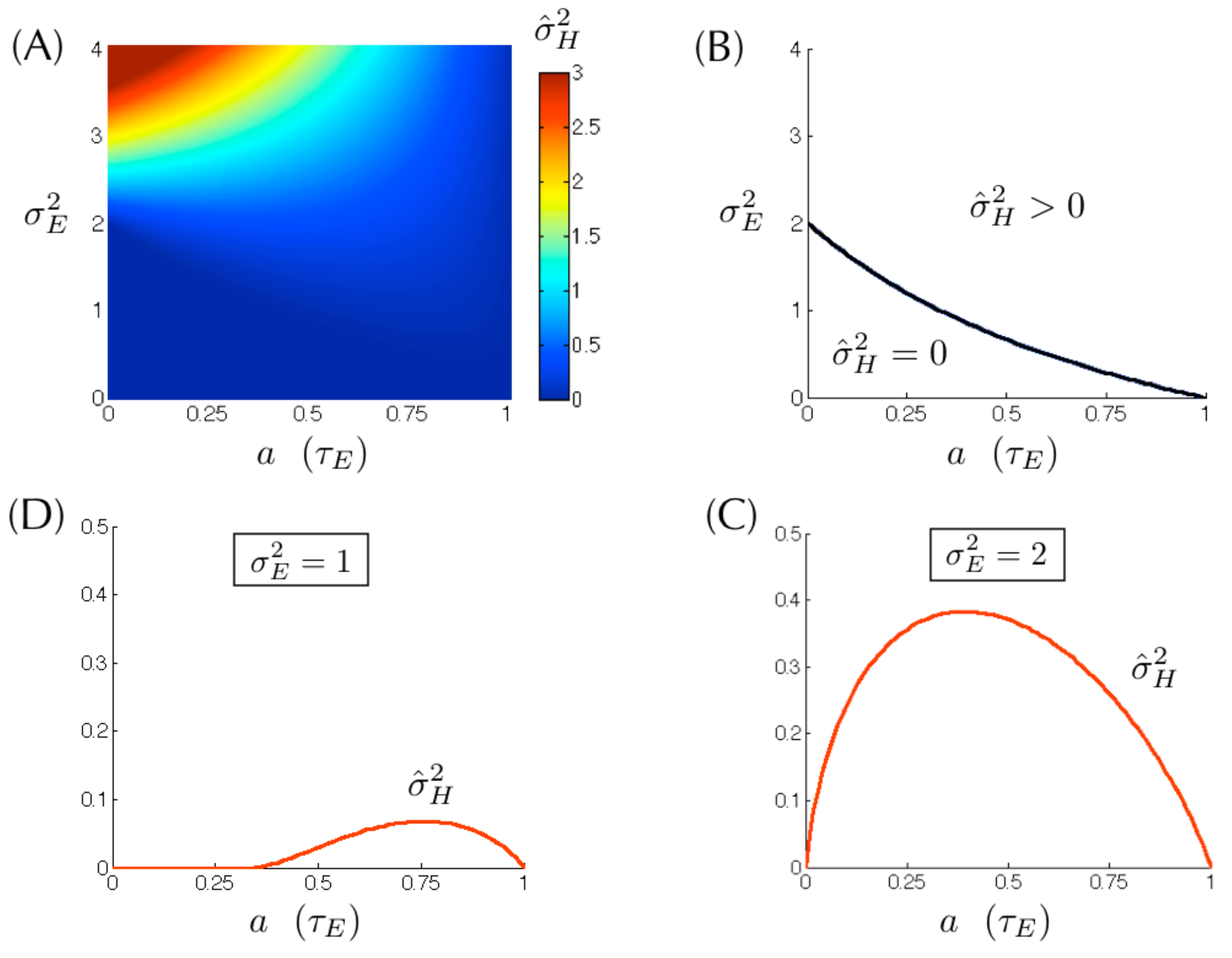}
\caption{Optimal degree of variation for the simple model of Sec.~\ref{sec:simple} -- {\bf (A)} Optimal value of $\sigma_H^2$ as a function of the environmental variables $(a,\sigma_E^2)$. {\bf (B)} Curve delimitating the environments in which $\hat\sigma_H^2=0$ is optimal. {\bf (C-D)} Same results presented for two fixed values of $\sigma_E^2$, $\sigma_E^2=1$ and $\sigma_E^2=2$.}\label{fig:simple}
\end{center}
\end{figure}

The results of an optimization of $\Lambda$ over $\sigma_H^2$ for the model of Sec.~\ref{sec:simple} are presented in Fig.~S\ref{fig:simple}. They show an optimal value $\hat\sigma_H^2$ for the introduction of new variations that depends on the environmental parameters $(a,\sigma_E^2)$, with no new variations ($\hat\sigma_H^2=0$) being beneficial for weakly correlated and weakly stochastic environments.

\subsection{Model of Sec.~\ref{sec:wherevar}}

The model of Sec.~\ref{sec:wherevar} is defined from the general model by taking $\l=1$, $\theta=1$, $\k=0$, $\omega=0$, $\rho=0$ and $\sq_S=1$. In this limit, Eq.~\eqref{eq:general} becomes

\beq\label{eq:L2}
\Lambda=r_{\rm max}+\frac{1}{2}\ln\frac{\alpha}{1+\sigma_D^2}-\frac{(1-a)\alpha}{(1+\alpha)(1-a\alpha)}\frac{\sigma_E^2}{1+\sigma_D^2},
\eeq
with
\beq
\alpha=\frac{2}{2+\tilde\sigma_H^2+\left(\tilde\sigma_H^2(\tilde\sigma_H^2+4)\right)^{1/2}},\qquad  \tilde\sigma_H^2=\frac{\sigma_H^2}{1+\sigma_D^2}.
\eeq

Instead of optimizing $\L$ over $(\sigma_H^2,\sigma_D^2)\in[0,\infty[\times [0,\infty[$, we can equivalently but more conveniently optimize
\beq
\tilde\L=r_{\rm max}+\frac{1}{2}\ln\frac{\alpha}{1+\sigma_D^2}-\frac{(1-a)\alpha}{(1+\alpha)(1-a\alpha)}\frac{\sigma_E^2}{1+\sigma_D^2},
\eeq
over $(\alpha,\sigma_D^2)\in[0, 1]\times[0,\infty[$, where $\alpha$ is treated independently of $\sigma_D^2$: the two optimizations must give the same optimal value, $\max_{\sigma_H^2,\sigma_D^2}\Lambda(\sigma_H^2,\sigma_D^2)=\max_{\alpha,\sigma_D^2}\tilde\Lambda(\alpha,\sigma_D^2)$, and $\hat\sigma_D^2=0$ when optimizing $\L$ if and only if $\hat\sigma_D^2=0$ when optimizing $\tilde\L$, as well as $\hat\sigma_H^2=0$ when optimizing $\L$ if and only if $\hat\alpha=1$ when $\tilde\L$.\\

First, if we assume $\hat\sigma_D^2>0$, we must have $\partial\tilde\L/\partial\sigma_D^2=0$, which gives
\beq\label{eq:hatdelta}
1+\hat\sigma_D^2=2\frac{(1-a)\alpha}{(1+\alpha)(1-a\alpha)}\sigma_E^2.
\eeq
Assuming additionally $\hat\alpha<1$, we must have $\partial\tilde\L/\partial\alpha=0$, which gives
\beq
\hat \a=\frac{1-a}{2a}.
\eeq
For this solution with $(\hat\sigma_H^2>0,\hat\sigma_D^2>0)$ to hold, we need $(1-a)/(2a)<1$, i.e. $a>1/3$. We thus conclude that the boundary between $(\hat\sigma_H^2>0,\hat\sigma_D^2>0)$ and $(\hat\sigma_H^2=0,\hat\sigma_D^2>0)$ is given by $a=1/3$ (black line in Fig.~\ref{fig:FigQ1}C).\\

By the same argument of continuity, the boundary between $(\hat\sigma_H^2>0,\hat\sigma_D^2>0)$ and $(\hat\sigma_H^2>0,\hat\sigma_D^2=0)$ corresponds to Eq.~\eqref{eq:hatdelta} having for solution $\hat\sigma_D^2=0$, which gives $\sigma_E^2=((1+a)/(1-a))^2/4$ (blue curve in Fig.~\ref{fig:FigQ1}C).\\

If we now assume $\hat\sigma_D^2=0$ but $\hat\sigma_H^2>0$, we obtain an equation for $\hat\a$ from $\partial\tilde\L(\sigma_D^2=0)/\partial\alpha=0$. Substituting $\a=1$ in this equation gives the boundary between $(\hat\sigma_H^2>0,\hat\sigma_D^2=0)$ and $(\hat\sigma_H^2=0,\hat\sigma_D^2=0)$, namely $\sigma_E^2=2(1-a)/(1+a)$ (red curve in Fig.~\ref{fig:FigQ1}C).\\

Finally, assuming $\hat\sigma_H^2=0$ but $\hat\sigma_D^2>0$, we obtain an equation for $\hat\sigma_D^2$ from $\partial\tilde\L(\alpha=1)/\partial\sigma_D^2=0$, which gives $\hat\sigma_D^2=\sigma_E^2-1$. The boundary between $(\hat\sigma_H^2=0,\hat\sigma_D^2>0)$ and $(\hat\sigma_H^2=0,\hat\sigma_D^2=0)$ is therefore given by $\sigma_E^2=1$ (green line in Fig.~\ref{fig:FigQ1}C).\\

All together, we thus obtain the diagram presented in Fig.~\ref{fig:FigQ1}C, with boundaries described by Eq.~\eqref{eq:boundaries}.  The corresponding analytical expressions for the optimal values of $\sigma_H^2$ and $\sigma_D^2$ (shown in Fig.~\ref{fig:FigQ1}D-E) are:
\bea
& (\hat\sigma_H^2>0,\hat\sigma_D^2>0)  &:\quad \hat\sigma_H^2=\frac{2(1-a)(1-3a)^2}{a (1+a)^2}\sigma_E^2,\quad  \hat\sigma_D^2=4\left(\frac{1-a}{1+a}\right)^2\sigma_E^2-1,\\
& (\hat\sigma_H^2=0,\hat\sigma_D^2>0) &: \quad\hat\sigma_H^2=0,\quad \hat\sigma_D^2=\sigma_E^2-1,\\
& (\hat\sigma_H^2>0,\hat\sigma_D^2=0)   &: \quad\hat\sigma_H^2=\hat\varsigma_H^2(a,\sigma_E^2),\quad  \hat\sigma_D^2=0,
\eea
where $\hat\varsigma_H^2(a,\sigma_E^2)= (1-\a)^2/\a,$ with $\a$ solution of
\beq\label{eq:beta0}
\sigma_E^2=\frac{(1+\alpha)^2(1-a\alpha)^2}{2(1-a)\alpha(1+a\alpha^2)}.
\eeq

\subsection{Optimization over discrete value for the model of Sec.~\ref{sec:whengp}}

\begin{figure}[t]
\renewcommand{\figurename}{FIG. S}
\begin{center}
\includegraphics[width=.75\linewidth]{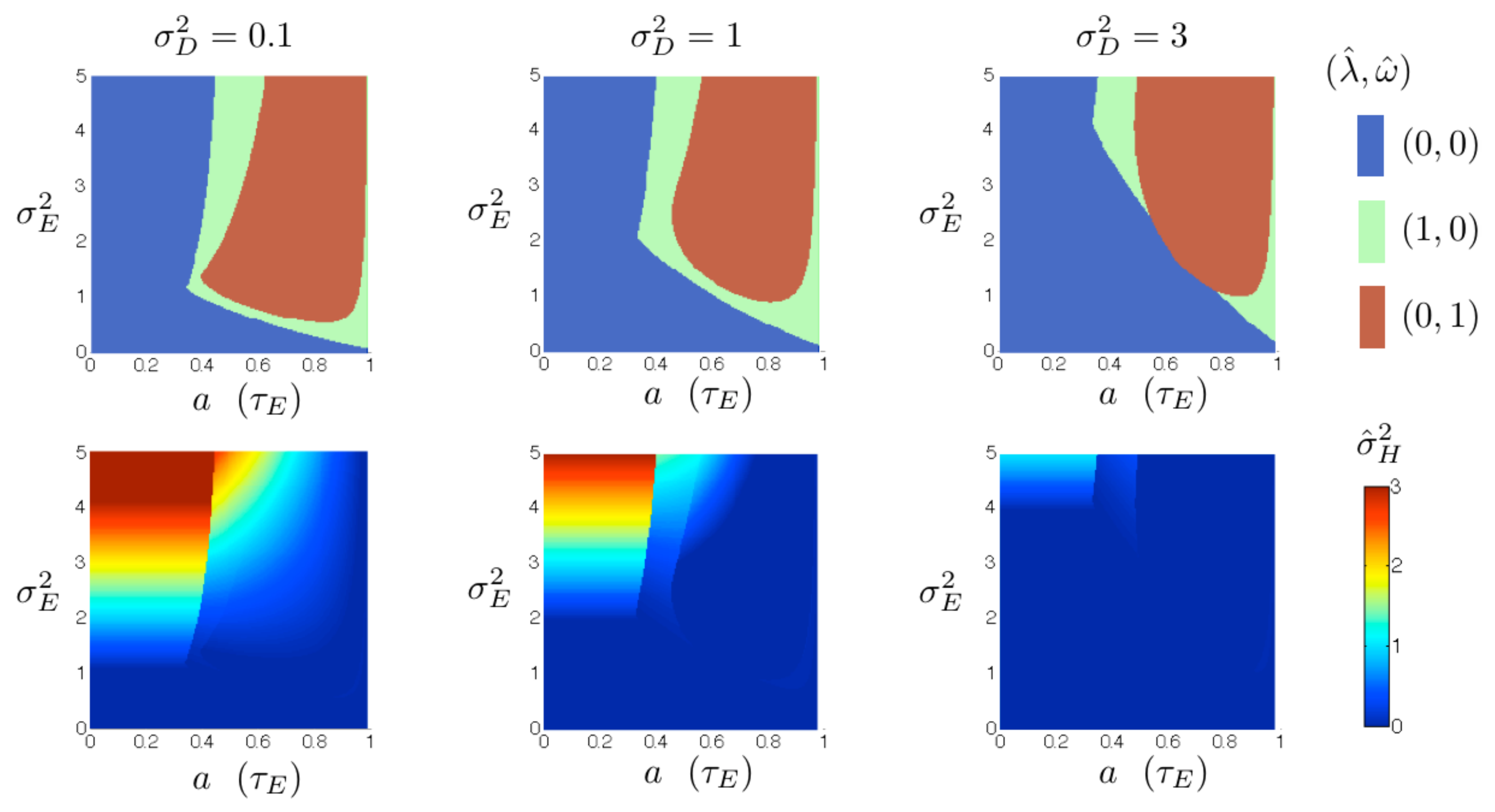}
\caption{When to separate phenotype and transmitted genotype -- For the model of Sec.~\ref{sec:whengp}, results of optimizations of $\Lambda$ over $(\lambda,\omega,\sigma_H^2)$ with $(\lambda,\omega)\in\{0,1\}^2$, for three fixed values of $\sigma_D^2$.}\label{fig:discrete}
\end{center}
\end{figure}

As an alternative to the optimization over $(\l,\omega,\sigma_H^2)$ for all $\l\geq 0$, $\omega\geq 0$ and $\sigma_H^2\geq 0$, we can restrain the optimization to $\l\in\{0,1\}$, $\omega\in\{0,1\}$ and $\sigma_H^2\geq 0$. The results, presented in Fig.~S\ref{fig:discrete}, are qualitatively similar.

\subsection{Optimal parameters for the models P and L of Sec.~\ref{sec:whereinto}}

\begin{figure}[t]
\renewcommand{\figurename}{FIG. S}
\begin{center}
\includegraphics[width=.7\linewidth]{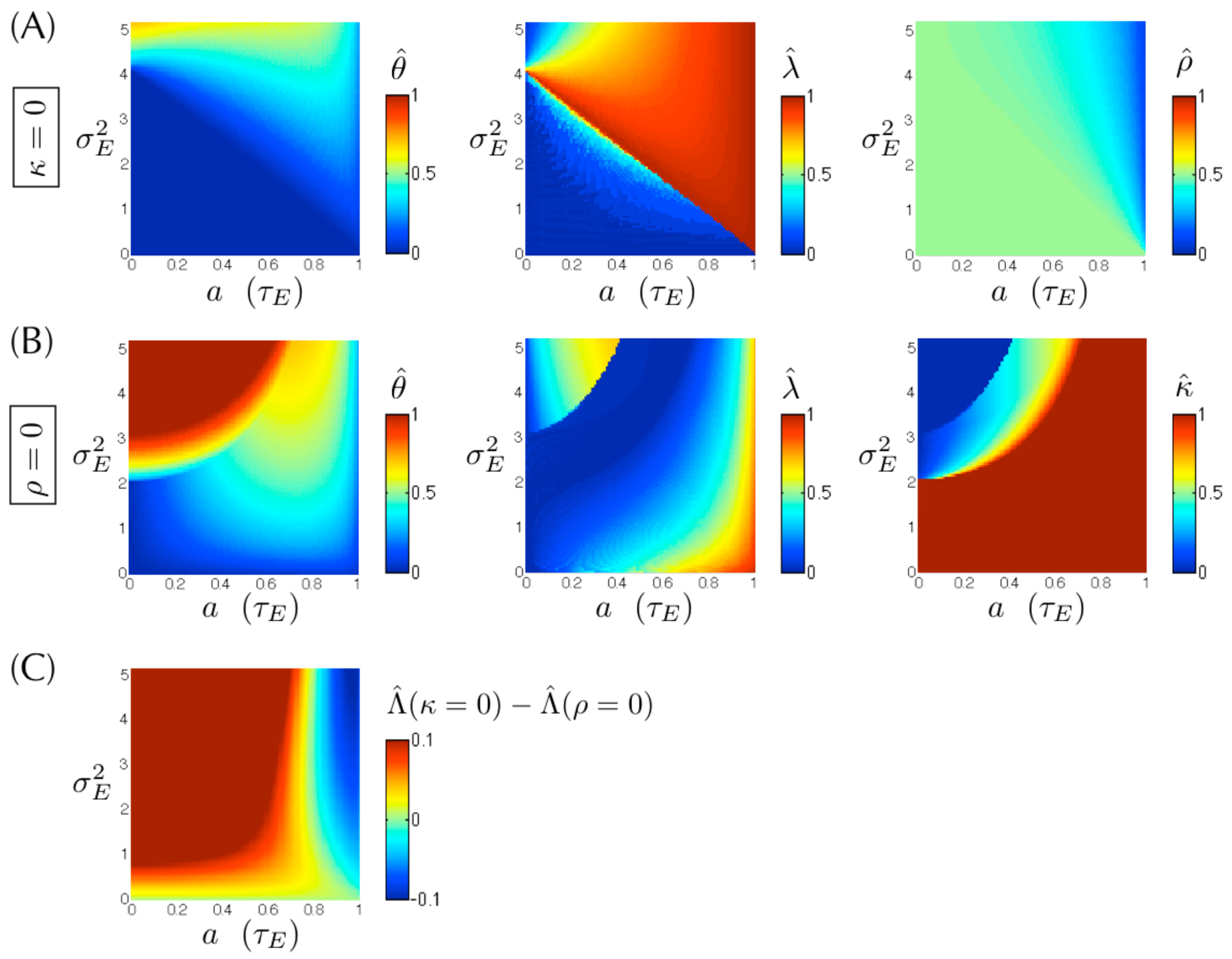}
\caption{Where to acquire information -- {\bf (A)} Model P: Phase diagram for the optimal values of $(\theta,\lambda,\rho)$ as a function of the environmental variables $(a,\sigma_E^2)$ for $\sigma_H^2=1,\sigma_D^2=1,\sigma_{Ip}^2=1$ and $\kappa=0$. {\bf (B)} Model L: Phase diagram for the optimal values of $(\theta,\lambda,\kappa)$ as a function of the environmental variables $(a,\sigma_E^2)$ for $\sigma_H^2=1,\sigma_D^2=1,\sigma_{I\ell}^2=1$  and $\rho=0$.  {\bf (C)} Difference between the optimal growth rates of model P, $\hat\Lambda(\kappa=0)$, and model L, $\hat\Lambda(\rho=0)$.}\label{fig:Q3S}
\end{center}
\end{figure}

Separate numerical optimizations of the models P and L for $\sigma_I^2=\sigma_H^2=\sigma_D^2=1$ lead to the results presented in Fig.~S\ref{fig:Q3S}. The comparison between the growth rate $\L$ of the two models is presented in Fig.~S\ref{fig:Q3S}C, which thus corresponds to the first panel of Fig.~\ref{fig:Q3}.

\section{Derivation of Price equation}

We consider here the simple model of Sec.~\ref{sec:simple}, described by Eqs~\eqref{eq:meanfield} and \eqref{eq:simplest}, i.e.,
\beq\label{eq:sm}
W_t\ n_{t+1}(\g')=\sum_{\xi}\xi \int\ud \g\ R(\xi|\g,x_t)\ H(\g'|\g)\ n_t(\g),
\eeq
where we do not make here any assumption on $R(\xi|\g,x_t)$ and $H(\g'|\g)$, besides the fact that they are stochastic kernels. We adopt the usual notation for writing Price equation, and denote the mean trait in the population, $m_t$, by $\langle \g\rangle_t$ and the mean "instantaneous fitness", $W_t$, by $\langle \xi\rangle_t$. The mean change of the trait between two generations, which is the focus of Price equation, is denoted by $\Delta_t\langle\g\rangle\equiv\langle\g\rangle_{t+1}-\langle\g\rangle_t$, and the covariance between the trait and the instantaneous fitness, which is the other central quantity, is defined by $\textrm{cov}_t(\xi,\g)=\langle\xi\g\rangle_t-\langle\xi\rangle_t\langle\g\rangle_t$, where the subscript $t$ accounts for the time-dependence of these quantities in our model.\\

Using the identity $\int\ud \g\ H(\g|\g')=1$, we first rewrite Eq.~\eqref{eq:sm} as
\beq
W_t\ n_{t+1}(\g')= \sum_{\xi}\xi\ R(\xi|\g',x_t) n_t(\g')+\sum_{\xi}\xi \int\ud \g\ \left[R(\xi|\g,x_t)\ H(\g'|\g)\ n_t(\g)-R(\xi|\g',x_t)\ H(\g|\g')\ n_t(\g')\right].
\eeq
After multiplying by $\g'$ and integrating over this variable, we thus obtain
\beq\label{eq:price1}
W_t\ \langle\g\rangle_{t+1}\equiv W_t\int\ud\g'\ \g'\ n_{t+1}(\g')=\langle \xi\g\rangle_t+\langle \xi\Delta\g\rangle_t,
\eeq
where $\Delta \g$, a function of $(\g,\xi,x_t)$, is defined by
\beq
\Delta \g\equiv\g\int\ud \g'\ \left[R(\xi|\g',x_t)\ H(\g|\g')\ n_t(\g')-R(\xi|\g,x_t)\ H(\g'|\g)\ n_t(\g)\right].
\eeq
Finally, subtracting $W_t \langle\g\rangle_t=\langle\xi\rangle_t\langle\g\rangle_t$ to both sides of Eq.~\eqref{eq:price1} leads to
\beq
\langle\xi\rangle_t\ \Delta_t\langle\g\rangle=\textrm{cov}_t(\xi,\g)+\langle \xi\Delta\g\rangle_t,
\eeq
which is the usual form of Price equation~\cite{Price:1970vw}. For our simple model, this formula decomposes evolutionary change in two terms: $\textrm{cov}_t(\xi,\g)$, the contribution from selection, and $\langle \xi\Delta\g\rangle_t$, the contribution from mutations, which vanishes when transmission is error-free, i.e., when $H(\g'|\g)=\delta(\g'-\g)$.\\

\end{document}